\documentclass[twocolumn]{aastex63}
\usepackage{amsmath}
\usepackage{amssymb}
\usepackage{threeparttable}
\usepackage{CJK}

\def\vel{\mathbf{v}}
\def\rs{r_{\rm s}}
\def\rp{r_{\rm p}}

\def\kappas{\kappa_{\rm s}}
\def\kappaa{\kappa_{\rm a}}
\def\kappap{\kappa_{\rm P}}
\def\rsi{r_{\rm SI}}
\def\Medd{\dot{M}_{\rm Edd}}
\def\mdot{\dot{M}}
\def\mdotout{\dot{M}_{\rm out}}
\def\rth{R_{\rm th}}

\begin{document}
\begin{CJK*}{UTF8}{gbsn}

\title{Pre-peak Emission in Tidal Disruption Events}
\author[0000-0003-2868-489X]{Xiaoshan Huang (黄小珊)}
\affiliation{California Institute of Technology, TAPIR, Mail Code 350-17, Pasadena, CA 91125, USA}
\affiliation{Department of Astronomy, University of Virginia, Charlottesville, VA 22904, USA}
\author[0000-0001-7488-4468]{Shane W. Davis}
\affiliation{Department of Astronomy, University of Virginia, Charlottesville, VA 22904, USA}
\author[0000-0002-2624-3399]{Yan-fei Jiang (姜燕飞)}
\affiliation{Center for Computational Astrophysics, Flatiron Institute, 162 Fifth Avenue, New York, NY 10010, USA}

\begin{abstract}
The rising part of a tidal disruption event light curve provides unique insight into early emission and the onset of accretion. Various mechanisms are proposed to explain the pre-peak emission, including shocks from debris interaction and reprocessing of disk emission. We study the pre-peak emission and its influence on the gas circularization by a series of gray radiation hydrodynamic simulations with varying black hole mass. We find that given a super-Eddington fallback rate of $10\Medd$, the stream-stream collision can occur multiple times and drive strong outflows of up to $9\Medd$. By dispersing gas to $\gtrsim100\rs$, the outflow can delay gas circularization and leads to sub-Eddington accretion rates during the first few stream-stream collisions. The stream-stream collision shock and circularization shock can sustain a luminosity of $\sim10^{44}\rm erg~s^{-1}$ for days. The luminosity is generally sub-Eddington and shows a weak correlation with accretion rate at early time. The outflow is optically thick, yielding a reprocessing layer with a size of $\sim10^{14}\rm cm$ and photospheric temperature of $\sim4\times10^{4}$K. 
\end{abstract}

\keywords{galaxies: ISM --- CR-magnetohydrodynamics --- ISM: jets and outflows ---method: numerical simulation}

\section{Introduction}\label{sec:introduction}

A star orbiting around a black hole can be tidally disrupted when its orbital pericenter is smaller than the tidal radius $R_{\rm T} = R_{\ast}(M_{\rm BH}/M_{\ast})^{1/3}$, where $R_{\ast}$ and $M_{\ast}$ are the radius and mass of the star, $M_{\rm BH}$ is the black hole mass. The accretion of the stellar debris can temporarily light up an inactive black hole, yielding a transient event that emits across a broad waveband \citep{rees1988tidal,ulmer1999flares,evans1989tidal}. These tidal disruption events (TDEs) now constitute a class of well-characterized transients \citep{gezari2021tidal}, becoming important probes to the accretion physics \citep{shiokawa2015general,piran2015disk,hayasaki2016circularization,liptai2019disc}, the properties of black holes such as their mass function \citep{rosswog2009tidal,metzger2016bright,mockler2019weighing} and spin distribution \citep{kasen2010optical,gafton2019tidal,mummery2024maximum}, as well as properties of their host galaxies \citep{french2020host,hammerstein2021tidal,somalwar2023vlass}. 

TDEs are mostly discovered in the optical/ultraviolet (UV) band \citep{van2011optical, hung2017revisiting, kochanek2017all, van2021seventeen,yao2023tidal,hammerstein2022final} or the X-ray band \citep{grupe1999rx,lin2011discovery,saxton2012tidal,sazonov2021first}. Early results suggested TDEs roughly formed two classes, with a single event preferentially emitting in either the optical/UV or X-ray bands, but rarely in both simultaneously. 

One theoretical interpretation for these two classes involves super-Eddington accretion leading to a puffed up, optically thick disk and outflow combined with a corona of hot electrons in the funnel region near the pole \citep{dai2018unified,thomsen2022dynamical}. The exclusive optical/UV or X-ray emission is then determined by the observers' viewing angle: the optical/UV events are identified with lines of sight through the disk/outflow that blocks the inner accretion region, and the X-ray events are from angles receiving emission directly from the corona region. In other models, accretion generates high energy photons, while the optical/UV emission is produced by other dynamical processes. For example, the collision shock due to the debris stream self-intersection (stream-stream collision) \citep{piran2015disk,jiang2016prompt} and the dynamical shocks from the gas around the black hole prior to accretion disk formation \citep{ryu2023shocks,steinberg2022origins}. Alternatively, after the accretion flow is established, the lower energy emission could arise from reprocessing by collision-induced outflow \citep{lu2020self, bonnerot2021first, huang2023bright}, optical-thick disk wind \citep{lodato2011multiband, miller2015disk, metzger2016bright, kara2018ultrafast} or a quasi-spherical, weakly bound envelope \citep{strubbe2009optical,roth2016x,metzger2022cooling,price2024eddington}.

Recent observations reveal a more diverse relationship between optical/UV and X-ray emission, complicating the earlier picture, where the lower energy and higher energy emission were largely exclusive to each other. \citet{malyali2023transient} report an example of X-ray detection roughly 14 days prior to its optical peak. \citet{guolo2023systematic} study the X-ray emission for optically selected samples from Zwicky Transient Facility. They show that the X-ray emission can emerge in different stages through out the optical light curve, and show diverse temporal evolution in addition to power-law decay. The ratio between black body optical luminosity and X-ray luminosity spans a wide range, suggesting there might be different types of dynamical evolution in TDEs. TDEs are also associated with both early-time and late-time radio emission with a range of different behaviours \citep{alexander2020radio, goodwin2022at2019azh,cendes2023ubiquitous} and potential powering mechanism \citep{matsumoto2021radio, spaulding2022radio, bu2023radio, matsumoto2023synchrotron}. It is likely that some mechanisms that proposed by different theoretical models operate simultaneously during TDE evolution, giving rise to the observed multi-band emission.

The rising part of the light curve is relatively less understood and particularly interesting for understanding the underlying physical processes driving the multiband emission. It has been shown that the optical rising time is weakly correlated to the black hole mass, but more likely to be affected by photon diffusion timescale \citep{van2021seventeen}, suggesting reprocessing might play an important role in pre-peak light curve. \citet{wang2023radio, 2024arXiv240301686H} find that some TDEs can present a pre-cursor flare in optical light curve before the main peak, which might be powered by stream-stream collision. Recent simulation work suggests that after the initial debris stream dissipates energy via stream-stream collision, gas circularization might be inefficient or delayed. The dynamical shocks are instead the leading emission mechanisms in early time \citep{ryu2023shocks,steinberg2022origins}. Understanding the pre-peak dynamics is essential to determine early energy dissipation mechanisms, which can be significantly distinct from standard accretion dominated dynamics. 

In \citet{huang2023bright}, we study the collision between two idealized TDE debris streams to understand the outcome of stream-stream collision by local high resolution radiation hydrodynamic (RHD) simulations assuming zero black hole spin. We find that stream-stream collision can yield a luminosity $\sim10^{42-44}\rm erg~s^{-1}$ and drive strong outflow away from the orbital plane, favouring the picture of shock powered emission and reprocessing by optically thick outflow from stream-stream collision. In this paper, we extend the simulation domain to global region around the black hole and follow the stream from the first passage of pericenter to tens of orbital periods at the stream-stream collision radius. We include time-dependent calculation of radiation field with tabulated opacity, which allows us to better capture radiation's role in gas dynamics and energy dissipation. The RHD simulations adopt gray opacity, but we also show one Monte Carlo post-processing simulation and one multi-group RHD simulation to explore the potential spectral information.

In Section~\ref{sec:method_set-up}, we introduce the numerical method, simulation set-up and the choice of stream orbits. The gray RHD simulations are listed in Table~\ref{tab:sim_params} with key parameters. Section~\ref{sec:result} shows the detailed results. We focus on the effect of black hole mass $M_{\rm BH}$, which strongly affect the debris stream orbit for given stellar mass, radius, and penetration factor. Our fiducial run RSI50Edd10 is presented in Section~\ref{subsec:result_fiducial}, where the simulated system corresponds to the case of a solar type star disrupted by $8\times10^{6}M_{\odot}$ black hole. In Section~\ref{subsec:restult_mbh} we vary the black hole mass and compare the dynamical processes. We show a separate suite of simulations to understand the potential effects of resolution in RHD simulations in Section~\ref{subsec:result_resolution}, focusing on the pericenter stream structure. In Section~\ref{sec:discussion}, we discuss the pre-peak emission sources (Section~\ref{subsec:discussion_emissionsource}) and energy dissipation mechanisms (Section~\ref{subsec:discussion_dissipation}). We discuss the potential high energy emission regarding to the viewing-angle effect, and potential radio emission from unbound gas produced in the collision (Section~\ref{subsec:discussion_multiband}). We compare the results with previous studies in Section~\ref{subsec:discussion_previouswork}. 

\section{Simulation Set-up}\label{sec:method_set-up}

\subsection{Equations and Units}
We solve the following equations in Athena++ with the explicit radiation transfer module \citep{jiang2021implicit}
\begin{eqnarray}\label{eq:RHD}
&\frac{\partial\rho}{\partial t}+\nabla\cdot(\rho\vel)=0,\label{eq:gasdensity}\\
&\frac{\partial(\rho\vel)}{\partial t}+\nabla\cdot(\rho\vel\vel-\textsf{P}^{*})=-\mathbf{G}+\rho\mathbf{a}_{\rm grav}\label{eq:gasmom}\\
&\frac{\partial E}{\partial t}+\nabla\cdot[(E+P)\vel]=-cG^{0}+\rho\mathbf{a}_{\rm grav}\cdot\vel,\label{eq:gasenergy}\\
&\frac{\partial I}{\partial t} + c\textbf{n}\cdot\nabla I=cS_{I}\label{eq:RT}\\
&S_{I}\equiv\Gamma^{-3}[\rho(\kappas+\kappaa)(J_{0}-I_{0})\nonumber\\
&+\rho(\kappas+\kappap)\left(\frac{a_{R}T^{4}}{4\pi}-J_{0}\right)]\label{eq:intensitysource}\\
&cG^{0}\equiv 4\pi c\int S_{I}d\Omega\label{eq:radenergy}\\
&\mathbf{G}\equiv4\pi\int\mathbf{n}S_{I}d\Omega\label{eq:radpressure}
\end{eqnarray}
Equation~\ref{eq:gasdensity} - Equation~\ref{eq:gasenergy} are gas hydrodynamic equations, where $\rho$, $\vel$, $E=E_{g}+(1/2)\rho v^{2}$ are fluid density, velocity and total energy density. We assume ideal gas law that relates internal energy $E_{g}$ and pressure $P$ by $E_{g}=P/(\gamma-1)$. $\textsf{P}$ is the pressure tensor. In the right-hand side of Equation~\ref{eq:gasmom} and Equation~\ref{eq:gasenergy}, radiation couples with gas through the components of the radiation four force $-\mathbf{G}$ and $-cG^{0}$, which are the momentum and energy exchange between radiation and gas. We adopt the generalized Newtonian gravitational description from \citet{tejeda2013accurate}. The gravitational force and gravitational potential are $\rho\mathbf{a}_{\rm grav}$ and $\rho\mathbf{a}_{\rm grav}\cdot\vel$, we list detailed expressions in \citet{huang2023bright}.

Equation~\ref{eq:RT} - Equation~\ref{eq:radpressure} represent the frequency integrated radiation transfer equation and the corresponding source terms. In the Athena++ implementation, the specific intensities $I$ are first transformed to the fluid comoving frame, where the opacities and emissivities are approximated to be isotropic \citep{jiang2021implicit}. The radiation source terms $S_{I}$, $\mathbf{G}$ and $cG^{0}$ are evaluated and updated along with the comoving frame gas internal energy equation. The resulting source terms are integrated over frequency and angle and then transformed back to the lab frame. Here, $c$ is the light speed. $I_{0}$ and $J_{0}$ are the comoving frame intensity and mean intensity. $\textbf{n}$ is the unit vector corresponding to rays in discretized angular grids, with the same angular discretization in \citet{davis2012radiation}. In the simulations, we used $n_{\mu}=4$, resulting $n_{\rm oct}n_{\mu}(n_{\mu}+1)/2=80$ angles across $n_{\rm oct}=8$ octants.  $\Gamma=\Gamma(\textbf{n}, \vel)=\gamma(1-\textbf{n}\cdot\vel/c)$ is the scaled Lorentz factor, with $\gamma=1/\sqrt{1-(v/c)^{2}}$ is the Lorentz factor. Equation~\ref{eq:radenergy} and \ref{eq:radpressure} represent radiation's effect on gas energy and momentum. 
In the source term, $a_{R}$ is the radiation constant, $\kappas$ and $\kappaa$ are scattering opacity and Rosseland mean absorption opacity, $\kappap$ is the the Planck mean opacity. We use OPAL opacity \citep{iglesias1996updated} for $\kappaa$ and TOPS opacity for $\kappap$ \citep{osti_6369029} based on the local gas density and temperature, and assume $\kappas=0.34~\rm cm^{2}~g^{-1}$. In the code, we solve the unit-less equations with the scaling of density unit $\rho_{0}=10^{-10}\rm g~cm^{-3}$, velocity unit $v_{0}=0.005c$ and length unit $l_{0}=r_{\rm s}=2GM/c^2$, where $M$ is the black hole mass. In the rest of paper, we report unit-less quantities unless explicitly specified with c.g.s unit.

\subsection{Resolution, Initial and Boundary Condition}\label{subsec:method_init}
We set the $\theta$ direction boundary to be outflow, $\phi$ direction boundary condition is periodic. The $r$ direction boundaries are set to be single direction outflow for hydrodynamical variables, which copies all the values from the first active cells but set any velocity that enters the calculation domain to be zero. The radiation boundaries in the $r$ direction is ``vacuum'' radiation boundaries, which copies all the intensities with $\textbf{n}$ pointing outward, but sets all intensities with $\textbf{n}$ pointing inward to be zero.

The simulations are performed in spherical polar coordinate with logarithmic grid in r direction. The calculation domain for RSI50Edd10 are $(2.1\rs,~267.2\rs)\times(0,\pi)\times(0,2\pi)$. RSI20Edd10 have domain size $(1.1\rs,~140.0\rs)\times(0,\pi)\times(0,2\pi)$, RSI100Edd10 run has domain size $(2.9\rs,~377.2\rs)\times(0,\pi)\times(0,2\pi)$. The base resolution in each direction is $64\times32\times64$. We add five levels of adaptive mesh refinement (AMR) to follow the high density regions. The refinement gives pericenter resolution $\delta R\approx0.018, 0.021, 0.028\rs$ in RSI20, RSI50 and RSI100 runs, equivalent to $\delta R\sim0.7R_{\odot}$ in all three simulations. 

We model the debris stream as a thin, cold stream injected near the black hole with a fixed mass fallback rate and angular momentum. In each simulation, the stream is injected at $r=r_{\rm inj}$, $\phi=\phi_{\rm inj}$ and $\theta_{\rm inj}=\pi/2$. The stream temperature is assumed to be $T_{\rm inj}=10^{5}$K. Following similar argument in \citet{jiang2016prompt}, the kinetic energy density in the streams is about six orders of magnitude higher than the internal energy density or radiation energy density before stream-stream collision, so the exact temperature of stream is unlikely to affect the shock dynamics. We also compared a simulation similar as RSI20Edd10 but with $T_{\rm inj}=10^{4}$K, we did not find significant difference in the dynamics or thermodynamics. 

We set density injection rate $\rho_{\rm inj}$ and velocity $\textbf{v}_{\rm inj}=(v_{\rm r,inj}, ~0, ~v_{\rm \phi, inj})$ in the neighboring four cells in $r$ and $\phi$ directions. The density injection rate at these cells is set by $\rho_{\rm inj}=f_{\rm Edd}\dot{M}_{\rm Edd}/(\mathbf{v}_{\rm inj}\cdot\mathbf{A}_{\rm inj})$, where $\dot{M}_{\rm Edd}=40\pi GM_{\rm BH}/(C\kappas)$ is the Eddington accretion rate, $\textbf{A}_{\rm inj}$ is the total injection area, and $f_{\rm Edd}$ is the Eddington ratio. We fix $f_{\rm Edd}=10$ in this work to study the super-Eddington fallback rate with different black hole mass. For higher mass, spin-less black hole and solar type star, $\beta=1.0$ orbits is more likely to yield partial disruptions \citep{mockler2019weighing,law2020stellar}, $f_{\rm Edd}=10$ for might be an idealized assumption. Nevertheless, RSI20Edd10 provides an interesting comparison in the limit of strong apsidal precession. In simulations, $\textbf{A}_{\rm inj}$ varies as we change $r_{\rm inj}$ and $\phi_{\rm inj}$, so we compute $\textbf{v}_{inj}\cdot\textbf{A}_{\rm inj}$ numerically for precise calculation. 

We set injection velocity $(v_{\rm r,inj}, ~0, ~v_{\rm \phi, inj})$ and injection cells location $(r_{\rm inj}, \pi/2.0, \phi_{\rm inj})$ based on ballistic trajectories that are integrated according to equations of motion from \citet{tejeda2013accurate} for a spin-less black hole. We assume the orbit eccentricity $e_{\rm orb}$ as the eccentricity for most bound material estimated as \citet{dai2015soft}
\begin{equation}\label{eq:ecc_orb}
    e_{\rm orb}\approx 1-0.02\left(\frac{M_{*}}{M_{\odot}}\right)^{1/3}\left(\frac{M_{\rm BH}}{10^{6}M_{\odot}}\right)^{-1/3}\beta^{-1},
\end{equation}
where $\beta=r_{\rm T}/\rp$ is the orbit penetration factor. $r_{\rm T}=R_{*}(M_{\rm BH}/M_{*})^{1/3}$ is the tidal radius, and $\rp$ is the pericenter radius. The orbit integration initial velocity $(0.0, 0.0, v_{\phi, 0})$ and location $(r_{0}, \pi/2.0, 0.0)$ are corresponding Keplerian values $r_{0}=\rp(1+e_{\rm orb})/(1-e_{\rm orb})$ and $v_{\phi,0}=\sqrt{GM_{\rm BH}/r_{0}}$. 

\begin{figure}
    \centering
    \includegraphics[width=0.5\textwidth]{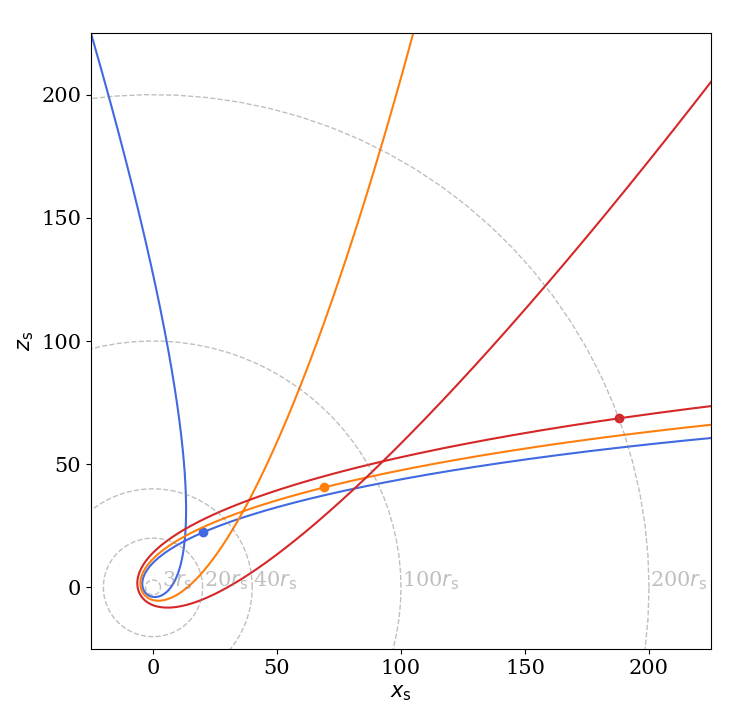}
    \caption{The integrated ballistic trajectories for black hole mass $M_{\rm BH}=10^{7},~8\times10^{6},~6\times10^{6}M_{\odot}$ (blue, orange and red), according to \citep{tejeda2013accurate}. The solid dots labels stream injection location in each simulation.}
    \label{fig:method_ballistic}
\end{figure}

We fix stellar mass and radius to be solar type star $M_{*}=M_{\odot}$ and $R_{*}=R_{\odot}$ in all simulations. Therefore, with fixed $\beta=1.0$, varying black hole mass $M_{\rm BH}$ changes stream orbit and stream-stream collision (stream self-intersection) location $\rsi$. Table~\ref{tab:sim_params} summarizes the $M_{\rm BH}$ adopted in each simulation. Figure~\ref{fig:method_ballistic} shows the ballistic orbits for $M_{\rm BH}=10^{7},~8\times10^{6},~6\times10^{6}M_{\odot}$ in this work. As we decrease the black hole mass, $\rsi$ moves further away from the black hole. 

For RSI50Edd10, $r_{\rm inj}=80.1\rs,~\phi_{\rm inj}=0.53$, $\textbf{v}_{\rm inj}=(-0.1c,~0,~0.03c)$, with $\rho_{\rm inj}=2.15\times10^{-6}\rm g~cm^{-3}$ for $f_{\rm Edd}=10.0$. For RSI20Edd10, $r_{\rm inj}=30.3\rs,~\phi_{\rm inj}=0.83$, $\textbf{v}_{\rm inj}=(-0.16c,~0,~0.07c)$, with $\rho_{\rm inj}=3.18\times10^{-5}\rm g~cm^{-3}$ for $f_{\rm Edd}=10.0$. For RSI100Edd10, $r_{\rm inj}=200.2\rs,~\phi_{\rm inj}=0.35$, $\textbf{v}_{\rm inj}=(-0.06c,~0,~0.01c)$, with $\rho_{\rm inj}=1.83\times10^{-7}\rm g~cm^{-3}$ for $f_{\rm Edd}=10.0$. The injection points are shown as the solid dots in Figure~\ref{fig:method_ballistic}. Before injecting stream, we set the initial background density and pressure to be low values of $\rho_{\rm init}=2\times10^{-8}$ and $P_{\rm init}=2\times10^{-12}$. We set density and pressure floor for the hydrodynamic Riemann solver to be $\rho_{\rm floor}=2\times10^{-8}$ and $P_{\rm floor}=2\times10^{-12}$. The temperature floor for radiation transfer module is set to be $T_{\rm floor}=2\times10^{-4}$.

\section{Results}\label{sec:result}

We perform three primary simulations RSI50Edd10, RSI20Edd10, RSI100Edd10, with a stream-stream collision radius (stream self-intersect radius) $r_{\rm SI}\sim 20, 50, 100 \rs$, respectively. We first describe the fiducial simulation RSI50Edd10 in Section~\ref{subsec:result_fiducial}, and then in Section~\ref{subsec:restult_mbh}, we describe results from RSI20Edd10 and RSI100Edd10 to study the effect of changing black hole mass, with other parameters fixed in Equation~\ref{eq:ecc_orb}.

\begin{table}
\caption{Summary of Simulation Parameters}
\label{tab:sim_params}
\begin{threeparttable}
\begin{tabular}{lccc}
\hline
Name & $M_{\rm BH}/10^{6}M_{\odot}$ & $\dot{M}/\dot{M}_{\rm Edd}$ & $r_{\rm SI}$ \\ 
\hline
RSI20Edd10 & 10.0 & 10.0  & 22.5$\rs$ \\
\hline
RSI50Edd10 & 8.0 & 10.0 & 47.7$\rs$ \\ 
\hline
RSI100Edd10 & 6.0 & 10.0 & 105.9$\rs$ \\ 
\hline
\end{tabular}
\caption{Simulation parameters including the black hole mass $M_{\rm BH}$ (the second column) normalized to $10^{6}M_{\odot}$, mass fallback rate $\dot{M}$ normalized to Eddington accretion rate of corresponding black hole mass (the third column) and stream-stream collision radius $\rsi$ (the fourth column). For $M_{\rm BH}=10.0, 8.0, 6.0\times10^{6}M_{\odot}$, the collision r adius $r_{\rm SI}=6.6\times10^{13}\rm cm,  ~1.1\times10^{14}\rm cm,  ~1.9\times10^{14}\rm cm$}
\end{threeparttable}
\end{table}

\subsection{Delayed Circularization due to collisionally induced outflow}\label{subsec:result_fiducial}

\begin{figure*}
    \centering
    \includegraphics[width=\textwidth]{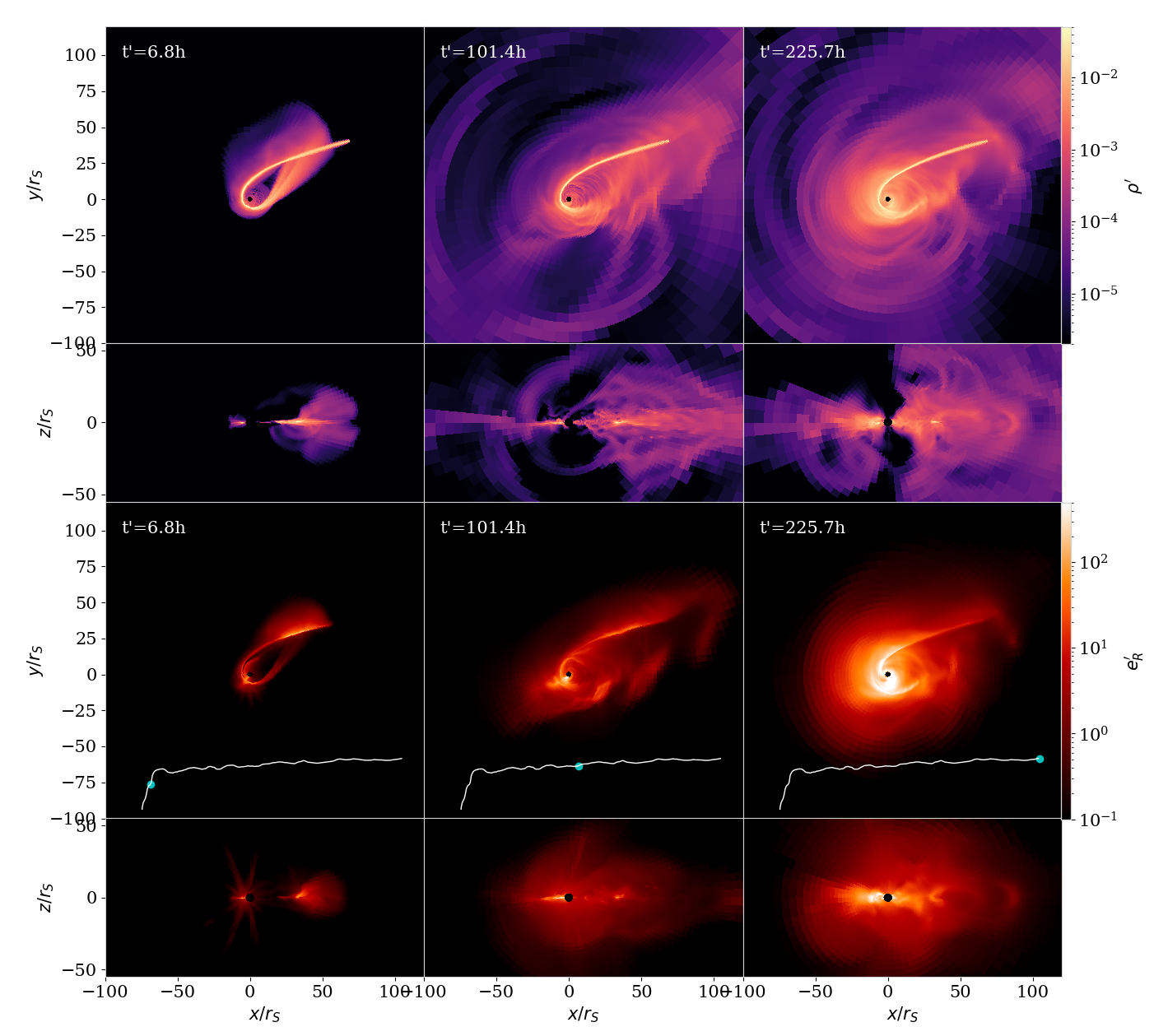}
    \caption{Gas density (the first and second row) and lab-frame radiation energy density (the third and fourth row) snapshots from RSI50Edd10. The three columns corresponds to $t=21.9,~94.6,~240.8$h, the white texts label the time since stream-stream collision in the c.g.s unit. The first and third row are quantities volume averaged over $\Delta\theta\sim5^{\circ}$ near $\theta=\pi/2$-plane, providing a ``face-on view''. The second and fourth plot shows volume average from $\Delta\phi\sim5^{\circ}$ near $\phi_{\rm inj}$ plane, providing a ``side view''. The white curve in the third row show the shape of estimated light curve in log scale (Figure~\ref{fig:mdotlum_mbh}), the blue dots labels time of current snapshot.}
    \label{fig:rsi50edd10_erdens}
\end{figure*}

In RSI50Edd10, as the returning stream passes pericenter, apsidal precession leads to stream-stream collision near $50\rs$ at the orbital plane. Figure~\ref{fig:rsi50edd10_erdens} shows a series of gas density and radiation energy density snapshots after the stream-stream collision. The first column shows that collision drives optically thick outflow in both the orbital plane and off the orbital plane. The post-shock gas is quickly heated and photons are thermalized, leading to enhanced radiation energy density near $\rsi$. 

\subsubsection{Strong Outflow Stage}
The initial stream-stream collision persists $t\approx18.0-27.1$h, during which the the returning stream impacts the fallback stream. The thin fallback stream is perturbed and thickened by the interaction. Around $t\approx27.1$h, the returning stream is disrupted and the stream-stream collision is briefly paused. The fallback stream reforms around $t\approx35.9$h, leading to another similar stream-stream collision around $t\approx 52.5$h. After each collision, small amount of bound, diffuse gas accumulates between the pericenter radius $\rp$ and collision radius $\rsi$. 

After the third collision, the returning stream expands significantly after passing the pericenter. The density contrast of the returning stream and the surrounding gas drops substantially. The stream structure is similar to the second column of Figure~\ref{fig:rsi50edd10_erdens}: a thin, high density fallback stream extends to pericenter, but disperses in the accumulated diffuse gas. The high eccentricity gas delivered by the expanded returning stream still interact with the thin fallback stream, but the interaction no longer resembles stream-stream collision at earlier times. A non-negligible fraction of gas starts to accumulate around the black hole, increasing the optical depth. The flow motion becomes more regular, circularizing near the black hole. The third column of Figure~\ref{fig:rsi50edd10_erdens} shows the next stage where gas accumulates near the black hole. We refer to this stage as ``circularization'' since the eccentricity of the gas is lower than the incoming stream, and decreases with time.  The eccentricity, however, is not zero and the flow deviates from Keplerian until the end of the simulation run. Gas density and radiation energy density between $\rsi$ and $\rp$ increases, the scale height of the circularizing gas also increases. The enhanced radiation energy density shifts from a concentration near the point of stream-stream collision to a more uniform distribution between $\rp$ and $\rsi$.

We observe multiple interactions between the fallback and the returning stream prior to flow circularization. Consistent with recent studies \citep{lu2020self,steinberg2022origins,huang2023bright,ryu2023shocks}, they produce prompt emission by converting kinetic energy to radiation. The first three encounters are the most violent ones, where the two streams both maintain sufficient density contrast with surrounding gas during the collision. 

\begin{figure}
    \centering
    \includegraphics[width=\linewidth]{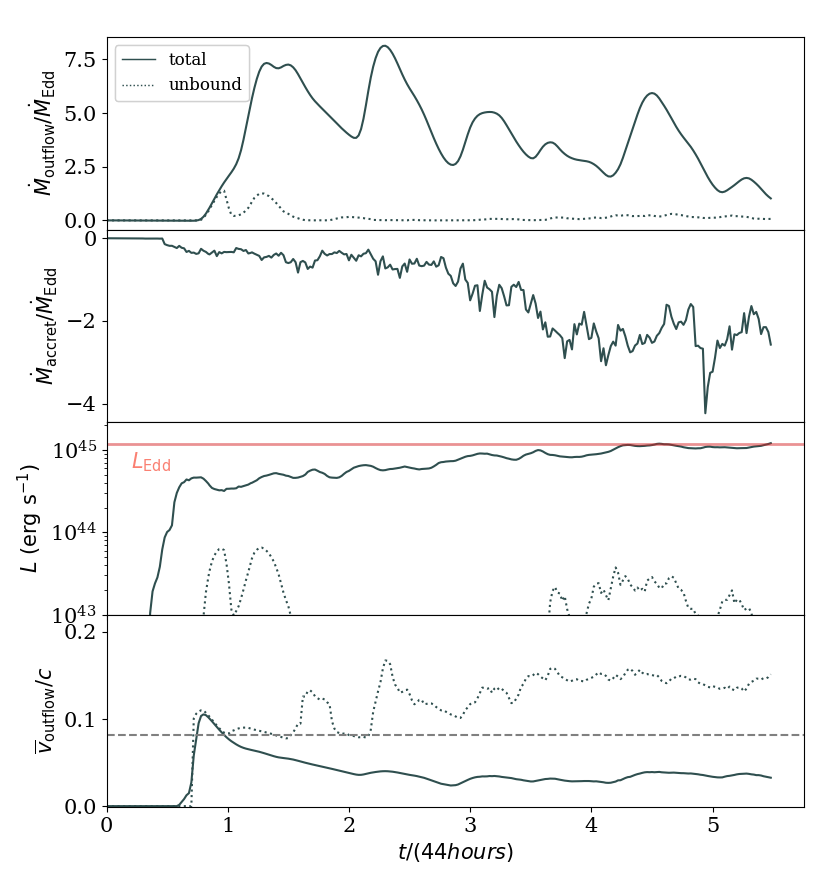}
    \caption{The outflow rate (the first row), the accretion rate (the second row), the luminosity (third row) and mass-weighted average outflow speed (the fourth row) of RSI50Edd10. In the first row, the solid line shows total mass outflow rate, normalized to $\Medd$. The dotted line shows the unbound outflow rate. In the third row, the dotted line shows the kinetic luminosity (Equation~\ref{eq:LKE_vout}) of carried by the unbound outflow, and the dotted line in the fourth row shows the mass-weighted average outflow speed for unbound gas. The horizontal dashed line labels the escape velocity at the measuring radius.}
    \label{fig:rsi50edd10_mdot}
\end{figure}

These major collisions drive significant outflow. Figure~\ref{fig:rsi50edd10_mdot} shows the total outflow rate measured at $r=150\rs$. (We find that the outflow rate is roughly constant once measured at $r\gtrsim90\rs$, so the specific choice of measurement radius is not important). Note that in our definition of outflow rate, gas is not necessarily unbound but could include bound gas that currently propagating away from black hole at large radius. The first three collisions happen around $t\approx21.9,~52.5,~87.6$h. The first outflow rate peak includes contributions from the first two collisions, giving peak outflow rate $\mdotout\approx7.5\Medd$ at $t\approx65.7, 96.3$h. 

Given that the mass flux carried by the injected stream is $\mdot=10\Medd$, the outflow rate $\mdotout\approx5-7.5\Medd$, which means that a significant fraction of gas is ejected as outflow instead of accumulating near the black hole. Therefore, accretion rate at these early time is suppressed. The second row in Figure~\ref{fig:rsi50edd10_mdot} shows the accretion rate measured at the innermost stable circular orbit (ISCO) $r_{\rm ISCO}=3\rs$. It is below $\Medd$ during the first three collisions, and only start to increase at $t\gtrsim127.0$h when the outflow rate drops. 

We show the mass-weighted average speed of outflow gas in the fourth row. The outflow in the first collision reaches average speed $\sim0.1c$, and drops to $\sim0.04c$ in the following two collisions. We found that in the second and third collision, the returning stream expands due to the interaction with the diffuse gas built up near the black hole from previous collisions. This slightly decreases the returning stream density but enhances the interaction area between the two streams, leading to comparable or larger outflow rate but slower outflow speed in the following two major collisions.

The peak outflow speed $\sim0.1c$ briefly exceeds the local escape velocity, indicative of unbound gas in the outflow. Similar to \citet{jiang2014global}, we define ``unbound gas'' as gas with $E_{\rm t}>0$, where $E_{\rm t}$ is:
\begin{equation}\label{eq:Et}
    E_{\rm t}=E_{\rm G}+\frac{\gamma P}{\gamma-1}+\frac{4E_{\rm rad}}{3}
\end{equation}
$E_{\rm G}$ is the sum of kinetic energy and gravitational potential. The second term is fluid enthalpy, and the third term assumes radiation as a fluid with effective adiabatic index of $\gamma=4/3$. We note that when radiation diffusion is present, $E_{\rm t}$ is not a conserved quantity. Nevertheless, it provides an analogue to Bernoulli number in an adiabatic fluid. 

\begin{figure}
    \centering
    \includegraphics[width=0.9\linewidth]{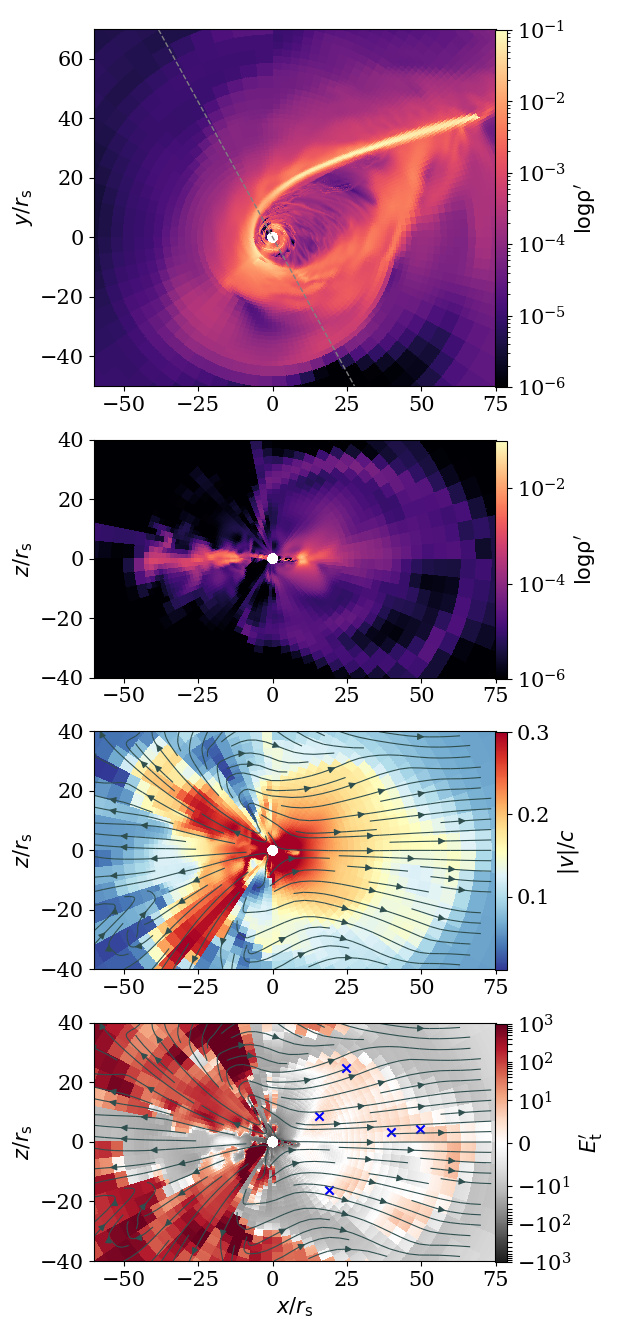}
    \caption{The first two column shows gas density snapshots from RSI50Edd10 at $t=56.1$h. The first row is a ``face-on'' view at the orbital plane, the plotted quantity is averaged over $\Delta\theta\sim5^{\circ}$ near $\theta=\pi/2$-plane. The second row is a ``side-view'' that cut through the gray dashed line in the first row, the plotted quantity is volume average from $\Delta\phi\sim5^{\circ}$ near $\phi=2.07$ plane. The third row is flow speed normalized to speed of light, the black solid lines shows the velocity streamlines at this projection. The fourth row shows $E_{\rm t}'$ (Equation~\ref{eq:Et}), positive value corresponds to unbound gas. The blue cross markers label selected sample points to extrapolate ballistic trajectories in Figure~\ref{fig:unbound_radio}.}
    \label{fig:rsi50edd10_unbound}
\end{figure}

We find that the fastest unbound gas is located primarily off the orbital plane. For example, Figure~\ref{fig:rsi50edd10_unbound} shows outflow in side-view slices at $\phi=0.66\pi$ during the second collision. The last rows shows the quantity $E_{\rm t}'$ defined Equation~\ref{eq:Et}, where positive $E_{\rm t}'$ near $x'\sim25-50$ indicated that this gas is unbound. At this time, they show bulk speed $\sim0.15c$ around $\sim30r_s$, decelerating as it propagates away from the black hole.

To estimate kinetic energy carried by the unbound gas, we define kinetic luminosity as:
\begin{equation}\label{eq:LKE_vout}
    L_{\rm KE}=\int\mathcal{F}_{\rm KE}\cdot\mathbf{dA}
\end{equation}
where $\mathcal{F}_{\rm KE}$ is kinetic energy flux through the sphere of radius $150\rs$, where we measure $L_{\rm kE}$. In the first, third and fourth row of Figure~\ref{fig:rsi50edd10_mdot}, we we show outflow rate, $L_{\rm KE}$ and mass-weighted average speed of unbound gas as dotted lines. The first two collisions create transient $\sim1.25\Medd$ unbound outflow, constituting roughly $\sim14\%$ of the fallback mass flux. Their average speed is relatively constant $\sim0.1c$. The unbound outflow rate drops to nearly zero after the second collision. The high average velocity in the late time are from small amount of unbound diffuse gas with density initialized at the numerical floor value.

In the third row, we compare the luminosity carried by radiation flux and the kinetic luminosity carried by unbound outflow. Around $t\sim21.9$h, the initial stream-stream collision leads to the prompt luminosity $L\approx8\times10^{44}\rm~erg~s^{-1}$, just below the Eddington luminosity $L_{\rm Edd}=1.2\times10^{45}\rm ~erg~s^{-1}$. The luminosity stays roughly constant, with only moderate fluctuations during the first three stream-stream collisions. The kinetic luminosity of unbound gas peaks during the first two collisions, which is $\sim10\%$ of luminosity carried by photons. At later time, when the outflow subsides, the unbound mass and kinetic energy flux are dominated by diffuse gas with typical densities close to the floor density of $\rho\lesssim10^{15}\rm g~cm^{-3}$.

The accretion rate increases simultaneously with the luminosity, albeit well below $\Medd$. We found that the collisions produce a small amount of more strongly-bound gas that is quickly accreted by the black hole, contributing to the accretion rate before outflow subsides. The lag between the rise of outflow rate and the rise of luminosity is consistent with the time that outflow propagates to the measurement radius, estimated as $\Delta t_{\rm out}\sim(r_{\rm 150\rs}-\rsi)/\overline{v}_{\rm outflow}\sim(150\rs-\rsi)/0.1c\approx21.9$h.  

\subsubsection{Gas Circularization Stage}

Around $t\gtrsim127.0$h, the thin, dense returning stream disperses shortly after the pericenter passage, well before it could collide with the incoming stream. The outflow rate drops and gas starts to accumulate near the black hole at a more rapid rate, the flow motion becomes less eccentric. The accretion rates grows from $\sim0.5\Medd$ to $\sim2.8\Medd$, the mass accumulated between $\rp$ and $\rsi$ increases by $\sim2.7$ times. The circularization yields relatively stable luminosity $\sim10^{45}\rm~erg~s^{-1}$, with slow increment over time. The scale height of circularizing flow also increases, as shown in the side-view density slice in third column of Figure~\ref{fig:rsi50edd10_erdens}. 

Although the flow eccentricity drops, we emphasize that it never becomes circular. It is not well-approximated by an axisymmetric, Keplerian disk. The fallback stream runs into the accumulating gas with lower angular momentum and orbital energy, forming localized shocks in the flow. A pair of weak spiral shock develops near the black hole, where radiation energy density is enhanced (see, for example, the second row of the third column in Figure~\ref{fig:rsi50edd10_erdens}). The spiral shock truncates where the dense fallback stream collides with the disk near $\rp$ and does not extends to outer region of circularizing flow. 

\begin{figure}
    \centering
    \includegraphics[width=\linewidth]{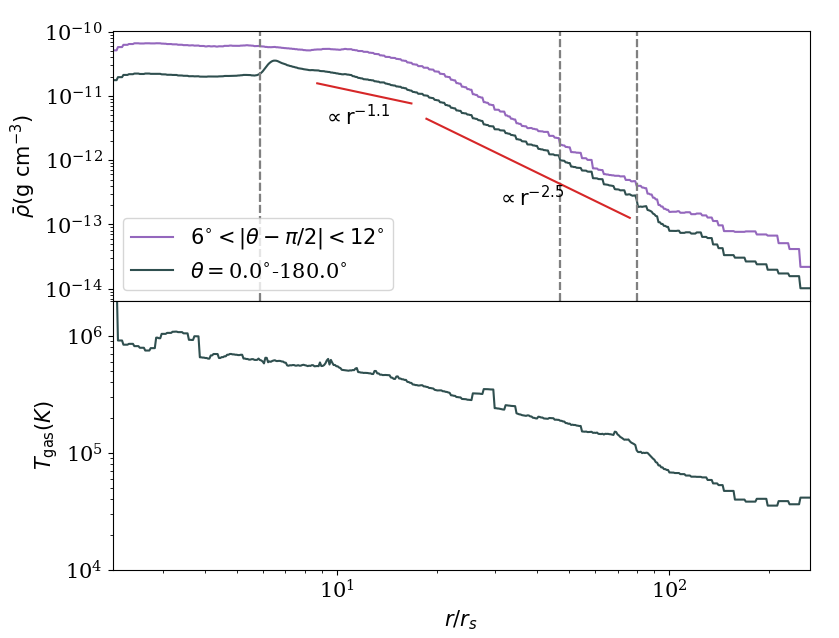}
    \caption{The average density (upper panel) and average gas temperature (lower panel) of RSI50Edd10 at $t=240$h. In the upper panel, the solid black line shows average gas density through spherical shells. The purple solid line shows the average gas density through a finite range of $\theta$ but excluding the orbital plane, $\pi/32<|\theta-\pi/2|<\pi/8$ near $\pi/2$. The three vertical dashed lines are tidal radius (also pericenter radius given $\beta=1.0$) $\rp$, stream-stream collision radius $\rsi$ and the injection radius $r_{\rm inj}$.}
    \label{fig:rsi50radial}
\end{figure}

Figure~\ref{fig:rsi50radial} shows radial profile of average gas density and temperature. The black solid lines are the averaged through spherical shells at each radius. The majority of gas accumulates between $\rp$ and $\rsi$ (the first and second vertical dashed lines), where the average gas density roughly follows power law. We can also describe the total density profile by two power law indexes, with a shallower inner region and steeper outer region separated around $\sim20\rs$. The best fitted indexes are $\rho\propto r^{-1.1}$ and $\rho\propto r^{-2.5}$. The inner region is more circularized while the outer region is affected by the stream-disk interaction and residual outflow. The black line in lower panel shows the average gas temperature.  Between $\rp$ and $\rsi$, the average gas temperature $T_{\rm gas}\sim10^{5}$K, and drops to $T_{\rm gas}\sim10^{4}$K towards the outer region. 

We also show the average density of a region with finite $\theta$ range but excluding the orbital plane, where the high density stream affects the average density. The off-orbital plane average density show more clear transition from the circularizing inner region $\rho\propto r^{-0.77}$ to the outer accumulating region $\rho\propto r^{-3}$. The average gas temperature profiles are similar to the spherically-averaged gas temperature and we do not plot them separately.

To quantify gas circularization, we define a scaled eccentricity following \citet{oyang2021investigating} as:
\begin{equation}
    \mathbf{e}=\frac{1}{GM_{\rm bh}}\mathbf{v}\times(\mathbf{r}\times\mathbf{v})-\mathbf{\hat{r}}
\end{equation}
In a circular Keplerian orbit, the magnitude of $|\mathbf{e}|=0$. With the gravity potential we adopt to approximate general relativistic precession, $|\mathbf{e}|$ is, however, non-zero for circular orbits at small radius. For example, $|\mathbf{e}|\approx4.47\times10^{-2},1.6\times10^{-2},1.2\times10^{-3}$ for $r=10\rs, 30\rs, 400\rs$. For the highly eccentric ballistic trajectory that we use to initialize the stream in RSI50Edd10 (Figure~\ref{fig:method_ballistic}), $|\mathbf{e}|$ drops from 1.0 at $r\gtrsim100\rs$ to its minimum $|\mathbf{e}|\approx0.6$ at $\rp$.

\begin{figure}
    \centering
    \includegraphics[width=\linewidth]{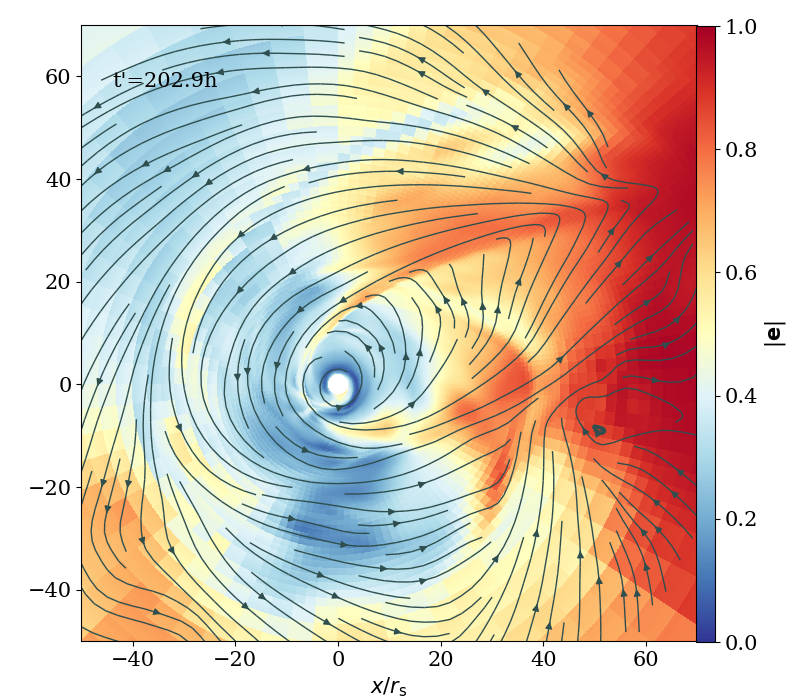}
    \caption{The magnitude of flow eccentricity (Equation~\ref{eq:ecc_orb}) in the orbital-plane at $t=240$h (RSI50Edd10). The dark gray lines are flow velocity streamlines. The gas with high eccentricity $|\mathbf{e}|\approx0.6-0.8$ roughly tracks the fallback stream. The circularizing gas $r<20\rs$ show lower eccentricity $|\mathbf{e}|\approx0.2$. The time in the label is time since the first stream-stream collision, equivalent to the last column in Figure~\ref{fig:rsi50edd10_erdens}.}
    \label{fig:rcoll50_ecc}
\end{figure}

In Figure~\ref{fig:rcoll50_ecc}, the inner circularizing region appears as the low eccentricity region near the black hole. The flow velocity stream lines are also more regular. The fallback stream's eccentricity is consistent with ballistic trajectory calculation. The kinks in velocity streamlines near $x\sim20-40,~y\sim20-40$ corresponds to the ``stream-disk'' shock between the high eccentricity fallback stream and the low eccentricity circularizing gas, where the local energy dissipation yields enhanced radiation energy density in Figure~\ref{fig:rsi50edd10_erdens}. 

\subsection{Varying the Black Hole Mass at Fixed $\beta$}\label{subsec:restult_mbh}

In RSI20Edd10 and RSI100Edd10, we fix the orbital penetration parameter $\beta$ (Equation~\ref{eq:ecc_orb}) and alter the ballistic trajectories by changing central black hole mass $M_{\rm BH}$. Figure~\ref{fig:method_ballistic} shows the corresponding orbits. When fixing the $\beta$, $M_{*}, R_{*}$, for spin-less black hole, increasing $M_{\rm BH}$ leads to more apsidal precession and smaller $\rsi$, so the stream-stream collision happens closer to the black hole \citep{dai2015soft, lu2020self}. Therefore, the velocity of streams before collision becomes larger, increasing the available kinetic energy to dissipate during the collision. Table~\ref{tab:sim_params} lists $\rsi$ in these runs, note that $\rsi$ is in unit of $\rs$ of corresponding $M_{\rm BH}$.

\begin{figure}
    \centering
    \includegraphics[width=\linewidth]{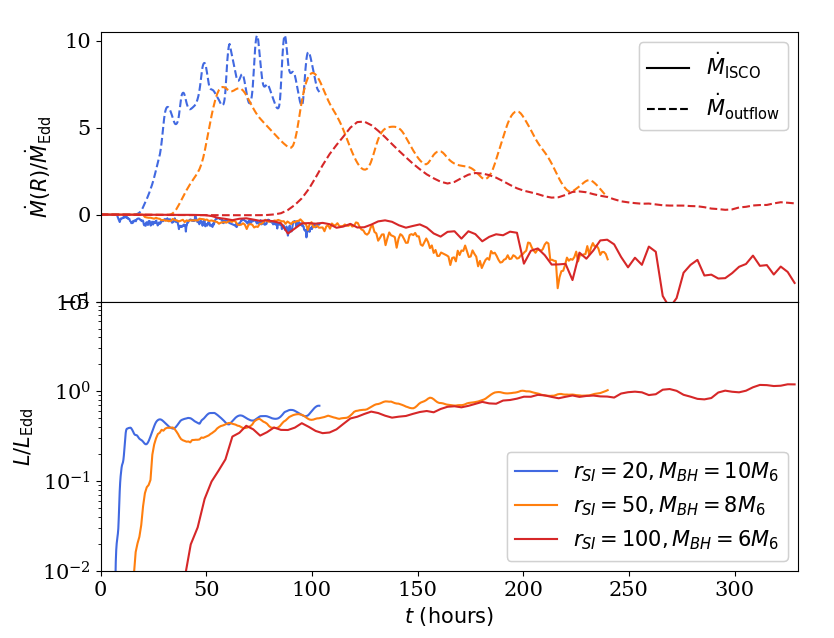}
    \caption{Upper panel: the outflow rates $\mdotout$ (dashed lines) and accretion rates (solid lines) in RSI50Edd10 (orange lines), RSI20Edd10 (blue lines) and RSI100Edd10 (red lines). The accretion rate is measured at $3\rs$ in all three runs. The outflow rate is measured over spherical shells at radius of $100\rsi$, $150\rsi$ and $300\rsi$ in RSI20Edd10, RSI50Edd10, RSI100Edd10. These are relatively outer radius in each simulation where the mass flux and luminosity is not sensitive to measuring radius. Lower panel: total luminosity $L$ carried by radiation flux measured over the same spherical shell as $\mdotout$, normalized to Eddington luminosity assuming $10\%$ efficiency. The prompt rise on $L/L_{\rm Edd}$ correspond to the emission from stream-stream collision. We note that such short rising timescale is related to the fixed $\mdot$ when injecting stream, the rising time in the simulations should not be interpreted as physical rising time for TDEs.}
    \label{fig:mdotlum_mbh}
\end{figure}

First, in RSI100Edd10, we consider the case where the stream-stream collision moves further away from black hole to $\rsi=102\rs$. The overall dynamics is similar to RSI50Edd10. The first stream-stream collision is the most prominent one, followed by weaker collisions between the expanded returning stream and thin fallback stream. Around $t\approx200.3$h, the fallback stream disperses near the pericenter, and gas begins to more rapidly accumulate near the black hole. 

Figure~\ref{fig:mdotlum_mbh} shows the outflow rate measured at $300\rs$ and accretion rate measured at $3\rs$. Compared with RSI50Edd10, the larger collision radius in RSI100Edd10 reduces the stream velocities before collision, leading to weaker interaction. The outflow rate peaks at $\mdotout\approx5\Medd$ around $t\approx137.0$h. Similar to RSI50Edd10, the first outflow peak covers the first two collisions. The outflow subsides around $t\approx6.1$ due to the significant expansion of returning stream, the average density near the black hole increases, and the flow eccentricity drops. Gas circularization near the black hole enhances the subsequent accretion rate. The lower panel of Figure~\ref{fig:mdotlum_mbh} shows that at late time, the accretion rate in RSI100Edd10 is slightly larger than RSI50Edd10 with an increasing trend.

The stream-stream collision again yields a prompt rise in the luminosity. The luminosity is slightly less variable and remains at a slightly lower quasi-steady-state value of $L\approx6\times10^{44}\rm erg~s^{-1}$, which is consistent with weaker collision. Overall, the luminosity is only weakly dependent on the location of the collision radius in these runs, and shows similar low fraction of Eddington luminosity despite the super-Eddington fallback rate.

In the strong outflow stage, the dynamics is modulated stream-stream collision at $\rsi$. In the circularization stage, the gas accumulation usually extends between pericenter radius $\rp$ and $\rsi$. So the orbital period $t_{\rm SI}\equiv \pi/\sqrt{2GM_{\rm BH}/r_{\rm SI}^{3}}$ at the $\rsi$ is a characteristic dynamical timescale. For example, $t_{\rm SI}=1.71\approx56$h in RSI100Edd10 and $t_{\rm SI}=0.51 \approx22$h in RSI50Edd10. We ran RSI50Edd10 and RSI100Edd10 for about $12t_{\rm SI}$ and $6t_{\rm SI}$. Hence, despite the longer physical time, RSI100Edd10 experiences fewer $t_{\rm SI}$ than RSI50Edd10. The circularization in RSI100Edd10 begins more sooner due to the weaker stream-stream collisions and outflow, allowing gas to accumulate near the black hole within less $t_{\rm SI}$.

\begin{figure}
    \centering
    \includegraphics[width=\linewidth]{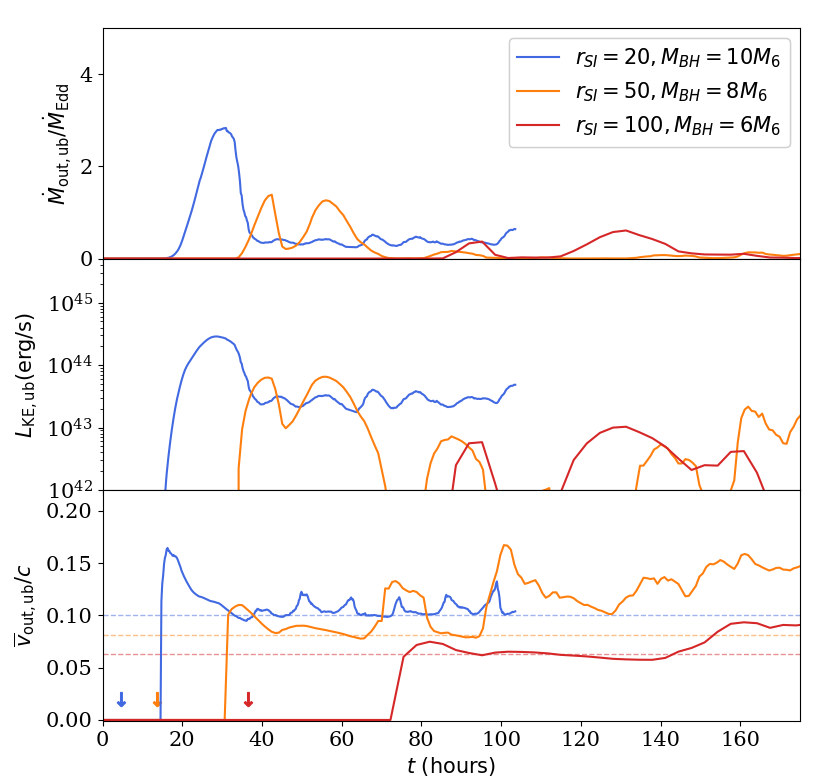}
    \caption{The first row: the unbound outflow rate in RSI50Edd10 (orange lines), RSI20Edd10 (blue lines) and RSI100Edd10 (red lines). The second and third row: kinetic luminosity $L_{\rm KE}$ (Equation~\ref{eq:LKE_vout}) and mass weighted outflow flow velocity for unbound gas. The outflow rate, kinetic luminosity and average velocity is measured over spherical shells at radius of $100\rsi$, $150\rsi$ and $300\rsi$ in RSI20Edd10, RSI50Edd10, RSI100Edd10. The arrows in the third row indicates the first stream-stream collision in the corresponding simulation. The horizontal dashed line in the third row shows the escape velocity at the measuring radius.}
    \label{fig:unbound_mbh}
\end{figure}

Weaker collision in RSI100Edd10 leads to less unbound gas in the outflow.  Figure~\ref{fig:unbound_mbh} shows the unbound outflow rate, kinetic luminosity and average velocity measured over the sphere shell at $300\rs$ for RSI100Edd10. The first three collisions are the primary source for the two unbound outflow rate peaks at $t=120.6, ~170.7$h, yielding a peak unbound mass flux $\dot{M}_{\rm out, ub}\approx0.6\Medd,~0.45\Medd$. The mass weighted average velocity indicates that the outflow is marginally unbound with velocity close to escape velocity. The arrows in Figure~\ref{fig:unbound_mbh} label the time of the first stream-stream collision. Due to the slower outflow and larger measurement radius at $300\rs$, the delay between the first stream-stream collision and the rise in the unbound outflow rate is longer than in run RSI50Edd100. The kinetic luminosity $L_{\rm KE}$ in run RSI100Edd10 is lower than in run RSI50Edd10. 

When we increase black hole mass to $M_{\rm BH}=10^{7}M_{\odot}$ in run RSI20Edd10, the apsidal precession is stronger and yields $\rsi\sim20\rs$. This implies larger encounter velocities and a stronger collision. Given the $\mdot=10\Medd$ that could be higher than physical fallback rate of solar type star, RSI20Edd10 provides a comparison study in the strong-collision limit. It shows a dynamical evolution that is different from the RSI50Edd10 and RSI100Edd10. 

\begin{figure*}
    \centering
    \includegraphics[width=\textwidth]{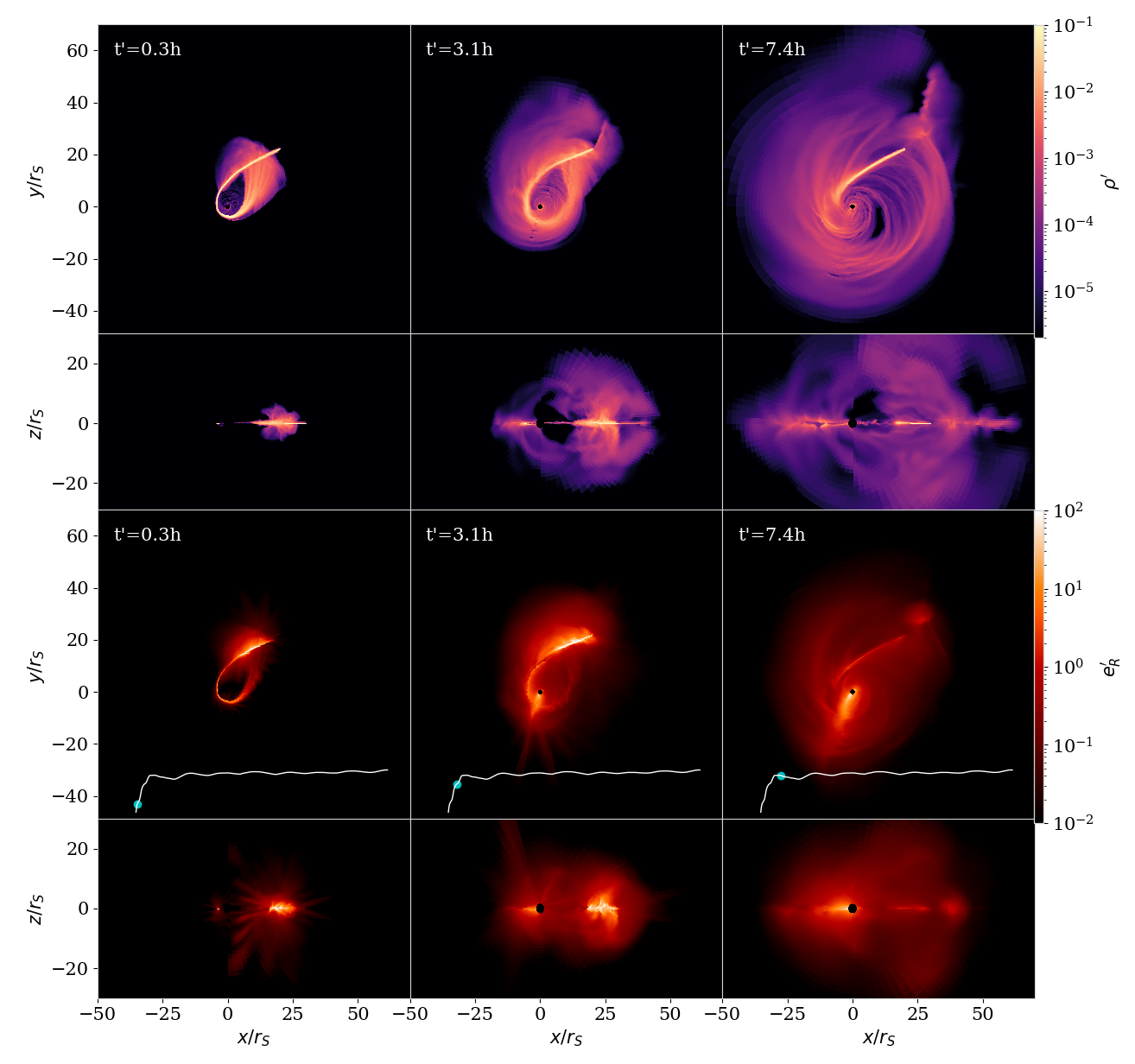}
    \caption{Gas density (the first and second row) and lab-frame radiation energy density (the third and fourth row) snapshots from Rsi20Edd10. The three columns corresponds to $t=7.6, 10.4, 14.7$h, the white texts label the time since stream-stream collision in the c.g.s unit. The first and third row are quantities volume averaged over $\Delta\theta\sim5^{\circ}$ near $\theta=\pi/2$-plane, the second and fourth plot shows volume average from $\Delta\phi\sim5^{\circ}$ near $\phi_{\rm inj}$ plane. The white curve in the third row show the shape of estimated light curve in log scale, the blue dots labels corresponding time of current snapshot.}
    \label{fig:rsi20edd10_erdens}
\end{figure*}

Figure~\ref{fig:rsi20edd10_erdens} shows process of the first stream-stream collision in RSI20Edd10. Similar to RSI50Edd10, the returning stream expands and collides with fallback stream, driving gas outflow and producing prompt emission. The radiation energy density is enhanced near $\rsi$. A prominent outflow is launched away from the orbital plane, an a larger fraction of the outflowing gas gains sufficient angular momentum to flow over the black hole instead of directly falling into the black hole from the polar region. When this gas returns orbital plane from above and below, they further collide and dissipate energy via shocks. As a result, a bright emitting column forms near the pericenter around $t=10.4-14.7$h, about $180^{\circ}$ from stream-stream collision site in the orbital plane.

The disrupted fallback stream reforms and leads to another stream-stream collision. Due to the close collision radius, the $t_{\rm SI}=9$h is rather small. We ran the simulation for about $12t_{\rm SI}$ to obtain a physical duration comparable to previous runs. Unlike RSI50Edd10 and RSI100Edd10, however, the system repeats similar streams-stream collision for about eight times and show little evidence of transition to a gas circularization stage where matter accumulates near the black hole. 

\begin{figure*}
    \centering
    \includegraphics[width=\textwidth]{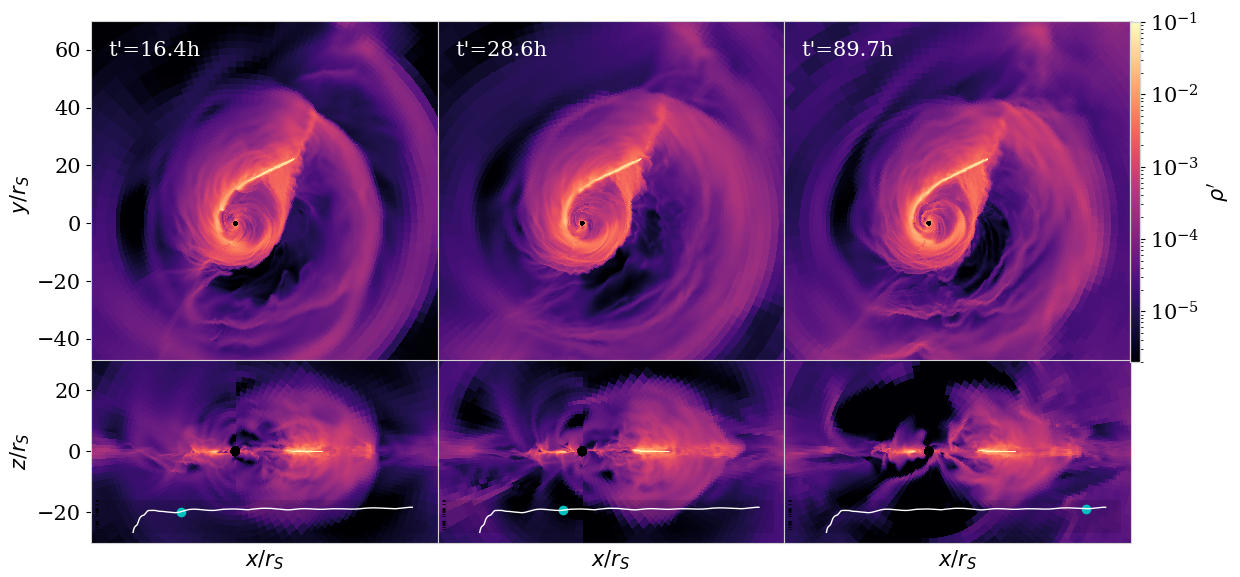}
    \caption{Gas density snapshots from Rsi20Edd10. The three columns corresponds to $t=23.5, 35.6, 95.8$h, the white texts label the time since stream-stream collision in the c.g.s unit. The first row are quantities volume averaged over $\Delta\theta\sim5^{\circ}$ near $\theta=\pi/2$-plane, the second row show volume average from $\Delta\phi\sim5^{\circ}$ near $\phi_{\rm inj}$ plane. The white curve in the third row show the shape of estimated light curve in log scale, the blue dots labels corresponding time of current snapshot.}
    \label{fig:rsi20edd10_repeats}
\end{figure*}

Figure~\ref{fig:rsi20edd10_repeats} compares the gas density at a similar dynamical stage during the second, third and eighth collision. At this moment in each collision, the returning stream already encounters the fallback stream and about to break the fallback stream. The average gas density around the black hole is increased compare to the first collision in the second column of Figure~\ref{fig:rsi20edd10_erdens}, but barely changes between these following collisions. 

Figure~\ref{fig:mdotlum_mbh} show the significant increase in the outflow rate during these repeating collisions. The $\mdotout$ of the first two collisions is comparable to the peak outflow rates in the other two runs with lower $M_{\rm BH}$ and larger $\rsi$, but the following collisions shows persistently higher outflow rates. The majority of the mass flux carried by the injected fallback stream is ejected as outflow. Consequently the accretion rate through $r_{\rm ISCO}$ is low due to lack of gas accumulation near the black hole.

Despite the persistence of episodic collisions, the luminosity shows only moderate variations. The periodicity of emission from stream-stream collision is smoothed over the time it takes photons to slowly diffuse out from optically thick region surrounding the collision point. The shocks produced by interaction of outflowing gas in between collisions also contributes to the emission. For example, some of the initially outflowing gas falls back to orbital plane and collides, yielding emissions by shock dissipation. The periodically-launched outflow also form multiple shells, the collision between the shells can also contributes to emission. These outflow interaction further reducing the luminosity difference between the active collision and stream reformation stages. 

The strong collisions in RSI20Edd10 creates more unbound outflow, even though these collision occur deeper in the potential. In Figure~\ref{fig:unbound_mbh}, the unbound $\mdotout$ peaks during the first two collisions around $3\Medd$, with average velocity $v_{\rm out,ub}\sim0.12c$. The unbound outflow rate drops below $\Medd$ in the following collisions but remains higher than in the other runs. The average velocity approaches to local escape velocity $\sim0.1c$. The persistent kinematic luminosity in the following collisions provides a potential source of kinetic energy from these marginally unbound gas. If this unbound gas can travel to large radius without significant dissipation, it may play an important role in early radio emission when shocking ambient gas, a possibility we discuss in Section~\ref{subsec:discussion_multiband}. 

We note that the periodic dynamics in RSI20Edd10 might be related to the assumption of a constant angular momentum and a rather higher fallback rate fixed at $\mdot=10\Medd$ in our injected stream. The effect of changing $\mdot$ and angular momentum distribution within debris stream will be a focus of future work. Nevertheless, the fallback rate for $M_{\rm BH}=10^{7}M_{\odot}$ is likely to be within similar order of magnitude for a few $t_{\rm SI}=8.8$h. The lack of gas accumulation in RSI20Edd10 suggests the potential scenario that the prominent mass loss from outflow pauses the formation of accretion flow. The three runs with different $\rsi$ show consistent trend that a stronger collision leads to longer delay of gas circularization. Increasing $\dot{M}$ of the fallback stream could also enhance the outflow because of larger kinetic energy flux carried by the streams. It is likely that a rising $\mdot$ at pre-peak times could produce non-linear mass accumulation near the black hole. 

\subsection{Resolution and Pericenter Stream Width}\label{subsec:result_resolution}

Near the pericenter, the gas ballistic trajectories with slightly different inclinations intersect near the pericenter. This effect results maximum vertical compression at the pericenter (perpendicular to the orbital plane), forming the ``nozzle shock''. The nozzle shock compression is commonly identified in previous simulations \citep{shiokawa2015general, hayasaki2016circularization, liptai2019disc, curd2021global, steinberg2022origins, ryu2023shocks, price2024eddington}. Depending on the black hole mass, pericenter radius and prescribed physical processes, it dissipates stream kinetic energy and leads to different levels of orbital-plane and vertical stream expansion.

Recent studies suggest that the stream can be compressed to be orders of magnitudes thinner near the pericenter compared to its original width \citep{bonnerot2022pericenter, bonnerot2022nozzle, coughlin2023dynamics}, making it numerically challenging to resolve the stream structure in proximity to the black hole. Other physical processes such as recombination might also impact stream width during the pericenter passage \citep{coughlin2023dynamics, steinberg2022origins}. In our simulations, the pericenter resolution for RSI20Edd10, RSI50Edd10 and RSI100Edd10 are $R\Delta\theta\approx0.018, 0.021, 0.028\rs\approx0.7R_{\odot}$. For the simulations reported in other sections, we set the maximum refinement level to five and ensure the pericenter is on highest refinement level. In this section, we use the orbit in RSI50Edd10 and compare five runs with maximum refinement level from three to seven. We ran the level three to six simulations to $t=32.8$h (except for RSI500Edd10 at level five), roughly right before the first stream-stream collision, the level seven run only ran to $t=13.1$h, when the stream just arrives pericenter.

We find that in the orbital plane, the stream expands after the first pericenter passage regardless of the resolution. Interestingly, the orbital-plane spreading is rather moderate in level three and four runs, where the returning stream roughly maintains its width before the pericenter compression. Increasing resolution to level five and above, however, leads to enhanced orbital-plane expansion. The returning stream expansion in level six ($R\Delta\theta\approx0.01\rs$) is similar to level five ($R\Delta\theta\approx0.018\rs$), with slightly reduced width. In other words, thinner streams are produced at lower resolution, but the orbital-plane stream width shows relatively weak dependence on resolution from five refinement level and above. We note that we do not reach stream width convergence even given the highest resolution, stream orbital-plane expansion can still be related to numerical diffusion. The convergence study in \citet{price2024eddington} also suggests orbital-plane expansion can be overestimated with insufficient resolution.

As the stream is compressed, energy is dissipated and much of it turns into radiation.  If the stream remains optically thick, and the diffusion time is long compared to the pericenter passage, significant radiation pressure is present to re-expand the stream as it leaves pericenter, and the evolution is approximately adiabatic. Alternatively, if the stream is sufficiently optically thin so that cooling is significant, there will be less radiation pressure available to re-expand the stream after pericenter passage. For physically likely parameters, the stream should remain sufficiently optically thick so that the diffusion time is much longer than pericenter passage and little radiation escapes \citep{bonnerot2022pericenter, coughlin2023dynamics}. 

When injecting the stream and before the stream experiences strong tidal compression, an important length scale to resolve is the vertical optical depth of the stream. The transition of horizontal expansion from level four to level five is reasoned as follow: similar to \citep{huang2023bright}, we keep $\dot{M}$ fixed when changing our resolution and inject the stream over a fixed width $H$ in both the $\theta$ and $\phi$ directions. The density $\rho\propto H^{-2}$, where $H$ is our injected stream width, which scales proportional to the resolution. The optical depth across the stream is roughly given by $\tau\sim\rho\kappa_{s}H\propto H^{-1}$.  Insufficient resolution leads to artificially larger $H$ and the density $\rho$ is underestimated by factor of $H^{2}$  consequently, the optical depth $\tau$ is underestimated by factor of $H$. Before the stream arrives pericenter, energy might be lost via radiation diffusion due to the artificially reduced optical depth. Equivalently, radiative loss is too efficient in low resolution cases, the thin width of returning steam is closer to the isothermal limit (for example, radiative efficient models in \citet{hayasaki2016circularization}. When increase resolution to resolve the optical depth, the tidal compression on the stream is more adiabatic instead of isothermal.

\begin{figure*}
    \centering
    \includegraphics[width=\textwidth]{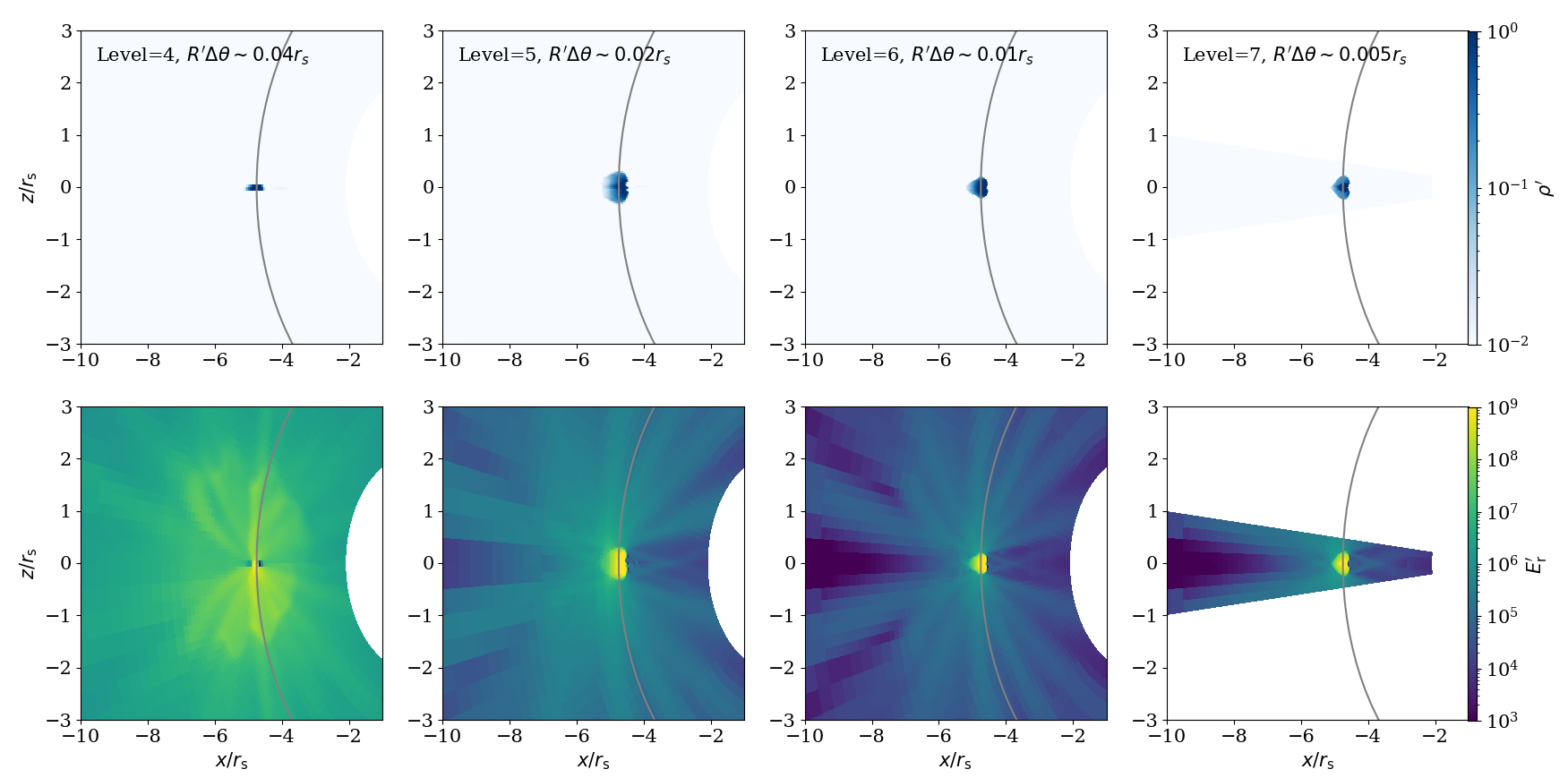}
    \caption{Snapshots of vertical stream density (upper panel) and lab-frame radiation energy density (lower panel) near pericenter. From left to right we show the stream vertical cross section for simulations with maximal refinement level from four to seven, the simulations reported in other sections are with level five refinement (the second column). In the highest refinement level, the simulation domain is the same as others, i.e. $\theta$ ranges from $0-2\pi$, we only read out and plot the data near orbital plane. The snapshot is cut at $r\approx4.75\rs$, $\phi=0.1\pi$ and $t=13.1$h. The orbital plane is at $z=0$. The solid gray curves label radius of $r=4.75$. }
    \label{fig:resedd10_erdens}
\end{figure*}

To elaborate the transition from level five and above, we show the ``side view'' density and radiation energy density in Figure~\ref{fig:resedd10_erdens} through a cross section at $\phi=0.1\pi$, near pericenter passage. In the set of comparison, we fix the fallback rate among all different resolutions. 

At this time, the stream is significantly compressed compare to the injected width. In the level four run (the first column), the stream is only a couple cells thick due to strong cooling before arriving pericenter. Insufficient resolution leads to an artificially large radiation flux. As a result, the radiation energy outside the dense stream is noticeably larger than in the other runs. After pericenter passage, there is little radiation pressure remaining to re-expand the stream and it remains thin. (Note that the level three run shows similar vertically collapsed stream as level four so we do not show it in Figure~\ref{fig:resedd10_erdens}.) 

We discuss the impact of resolution on the vertical structure of the streams further in Appendix~\ref{appendix:resolution}, but the upshot is that for the runs with maximum refinement level five and above, we start to resolve the optical depth and capture the vertical structure of the streams. Increasing resolution from level five to seven, the stream vertical height continues to decrease and the radiation energy outside the stream also decreases. 

In other words, we find that before the stream arrives pericenter and experiences strong compression, the radiative cooling decreases as the resolution increases, the stream becomes more optically thick.  The scale height of the stream in the vertical direction is still set by our resolution threshold and estimates suggest that we need at least an order-of-magnitude higher resolution to resolve the vertical height of the stream, and the corresponding horizontal expansion dependence on resolution from level seven and above is unclear. We find, however, that five levels of refinement reduces the flux to the point where the cooling is low enough that the radiation energy density within the stream remains high and the re-expansion of the stream is nearly independent of resolution. 

\section{Discussion}\label{sec:discussion}

\subsection{Pre-peak Emission from shocks: from stream-stream collision to circularization shock}\label{subsec:discussion_emissionsource}

In the conventional perspective, TDE emission originates from the accretion of stellar debris. For a significant fraction of TDEs, the late time light curve roughly follows the theoretical mass fallback rate, which sets the accretion rate.  While this may still be a reasonable picture of the later decay phase, we find that the rising time emission is more likely powered by various shocks instead of accretion. Figure~\ref{fig:mdotlum_mbh} shows in all three simulations, the measured luminosity is about a few $\sim10^{44}\rm erg~s^{-1}$ and rarely exceed Eddington luminosity. The accretion rate near $r_{\rm ISCO}$ is usually well below the mass fallback rate $10\Medd$. To quantify the relation between accretion and luminosity, we adopt the definition of radiation efficiency $\eta_{\rm out}=L/\mdot c^{2}$, where $\mdot$ is the accretion rate measured at ISCO.

\begin{figure}
    \centering
    \includegraphics[width=\linewidth]{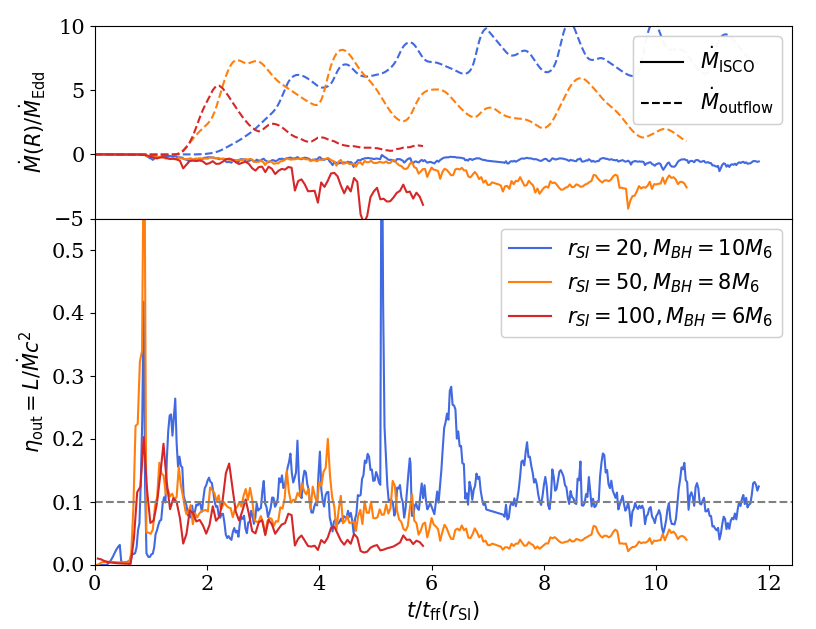}
    \caption{The upper panel shows outflow rates (dashed lines) and accretion rate (solid lines), the lower panel shows radiation efficiency $\eta$ for RSI20Edd10, RSI50Edd10, RSI100Edd10 (blue, orange and red lines). The time is normalized to the free-fall time at $\rsi$ to show the correlation with dynamics. The upper panel is equivalent to the upper panel of Figure~\ref{fig:mdotlum_mbh} except for different time normalization. The gray dashed line in the lower panel labels $10\%$ efficiency.}
    \label{fig:lumeff_mbh}
\end{figure}

We show the evolution of radiation efficiency $\eta$ in the lower panel of Figure~\ref{fig:lumeff_mbh}. We normalize time to the orbital period at stream-stream collision radius $t_{\rm SI}$, and in the upper panel, we also plot the outflow and accretion rate for comparison. The radiation efficiency evolution in the pre-peak time significantly deviates from the picture of a classic steady-state accretion disk. The efficiency $\eta$ fluctuates with time, and shows a complex relation to the accretion rate.

In RSI50Edd10 and RSI100Edd10, when the $\mdotout>5\Medd$ and the accretion rate is well below $\Medd$, the efficiency fluctuates around $\approx10\%$. When the outflow rate drops and accretion rate grows (roughly $6t_{\rm SI}$ for RSI50Edd10 and $3t_{\rm SI}$ for RSI100Edd10), the radiation efficiency tends to be lower and more stable at around $4\%$. The efficiency drops because the increased accretion rate but doesn't commensurately increase the luminosity, which is dominated by stream interactions.  RSI20Edd10 provides a limit where the outflow depletes the gas near black hole and halts circularization. Dominated by stream collisions, it shows a consistently low the accretion rate in $12t_{\rm SI}$, but the highest $\eta$ among the three runs at around $10\%$.

The lack of correlation between $\eta$ and accretion rate suggests that the pre-peak emission is not directly powered by accretion at small radius. We find that various shocks in the dynamical process are important sources for pre-peak emission. For RSI20Edd10, shown in Figure~\ref{fig:rsi20edd10_erdens}, the radiation energy density is enhanced by the stream-stream collision shock and the interaction between subsequent outflow. The luminosity from RSI20Edd10 is also roughly consistent with the comparable run in \citet{huang2023bright} and \citet{jiang2016prompt}, which are local studies focusing on the stream-stream collision itself without accretion.

In RSI50Edd10 and RSI100Edd10, prior to the circularization stage, the stream-stream collision is the primary source for emission. After gas accumulates near the black hole, the high radiation energy density regions shift from stream-stream collision point to nearer the black hole, as seen in Figure~\ref{fig:rsi50edd10_erdens}. Small spiral patterns develop near the ISCO, where the radiation energy density increases locally at the shock front. Another prominent high radiation energy density region traces the shock where the fallback stream is running into the circularizing gas. If such ``stream-disk'' interaction persists, the density contrast between the stream and circularizing gas might be an important factor determining the emission. These shocks can be modulated by the time-dependent mass fallback rate, and understanding their effect will be important for future work.

\subsection{Energy Dissipation in Post-shock Flow}\label{subsec:discussion_dissipation}

In RSI50Edd10 and RSI100Edd10, gas accumulates near the black hole after the outflow subsides, increasing the average radiation energy density between the tidal radius $r_{\rm T}=\rp$ and the stream-stream collision radius $\rsi$. Figure~\ref{fig:rcoll50_ecc} shows a snapshot of eccentricity magnitude (Equation~\ref{eq:ecc_orb}) in the orbital plane, where $|\mathbf{e}|$ in regions between $\rp$ and $\rsi$ is reduced compare to original ballistic eccentricity. 

\begin{figure}
    \centering
    \includegraphics[width=\linewidth]{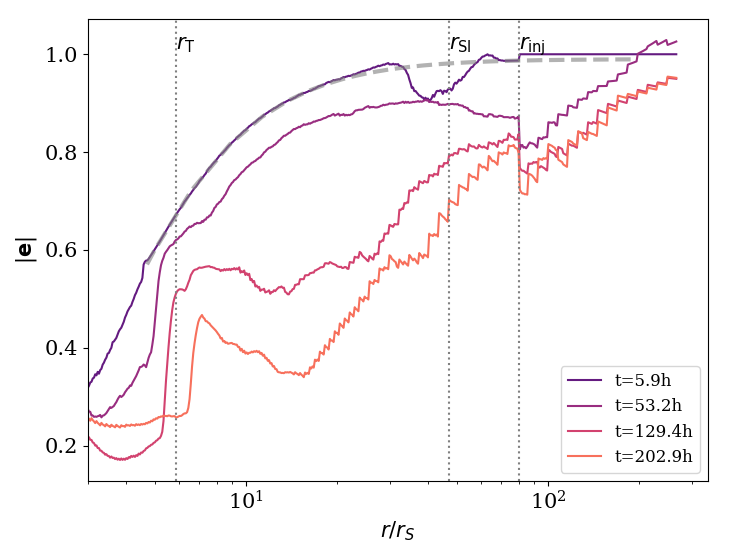}
    \caption{Eccentricity magnitude $|\mathbf{e}|$ (Equation~\ref{eq:ecc_orb}) averaged in $\theta$ and $\phi$ direction in RSI50Edd10. The gray dashed curve shows $|\mathbf{e}|$ for the initial stream ballistic trajectory. For the general Newtonian potential we adopt, $|\mathbf{e}|$ drops to $\approx0.6$ towards to the pericenter. The different colors of solid curves shows average $|\mathbf{e}|$ in different times (since the first collision). The vertical dotted lines labels three radius: the tidal radius $r_{\rm T}=\rp$, the stream-stream collision radius $\rsi$ and the radius where we inject the stream $r_{\rm inj}$.}
    \label{fig:radial_ecc_rsi50}
\end{figure}

We show $|\mathbf{e}|$ averaged in $\theta$ and $\phi$ direction in Figure~\ref{fig:radial_ecc_rsi50}. The gray dashed line is the $\theta$ and $\phi$ direction averaged eccentricity for the ballistic trajectory of injected stream. The eccentricity is close to one at the injection radius. Due to the non-Newtonian potential, it drops to roughly $0.6$ at the pericenter. The non-Newtonian potential also slightly shifts the pericenter radius from the tidal radius even though we assumed $\beta=1.0$. At $t=6$h, just after the stream-stream collision begins, the stream is only perturbed at near $\rsi$ so the majority of radius still follows the eccentricity of the ballistic trajectory. The eccentricity drops slightly in the strong outflow stage ($t=53$h), and only later in the gas circularization stage ($t=129,~202$h) does the eccentricity significantly decreases between $r_{\rm T}$ and $\rsi$. 

The eccentricity decrement suggest energy dissipation is associated with the outflow launching and gas circularization. The radiation production in the system provides a perspective to quantify the dissipation of total gas energy, including kinetic energy, gravitational potential and the relatively small internal energy. Recent work has investigated the ``circularization efficiency'' in terms of the ratio between radiative loss rate and the orbital energy required to circularize all the supplied gas around $2\rp$ \citep{steinberg2022origins,ryu2023shocks}. The circularization efficiency presented in early time ranges from $\lesssim5\%$ to $\sim30\%$ from previous works depending on dynamical system configuration.

We compare the radiative loss in all the simulations and found contrasting results for RSI50Edd10 and RSI100Edd10, which indicate a circularizing flow, and RSI20Edd10, which lacks evidence for significant gas circularization. We estimate the radiation loss $e_{\rm loss}(r)$ by normalizing the total radiation energy $E_{\rm rad}$ to the total circularization energy $E_{\rm circ}=GM_{\rm BH}M_{r}/2r$ within radius $r$:
\begin{equation}\label{eq:eloss}
    e_{\rm loss}(r) = \frac{E_{\rm rad}(<r)}{M(<r)(GM_{\rm BH}/2r)}
\end{equation}

\begin{figure}
    \centering
    \includegraphics[width=\linewidth]{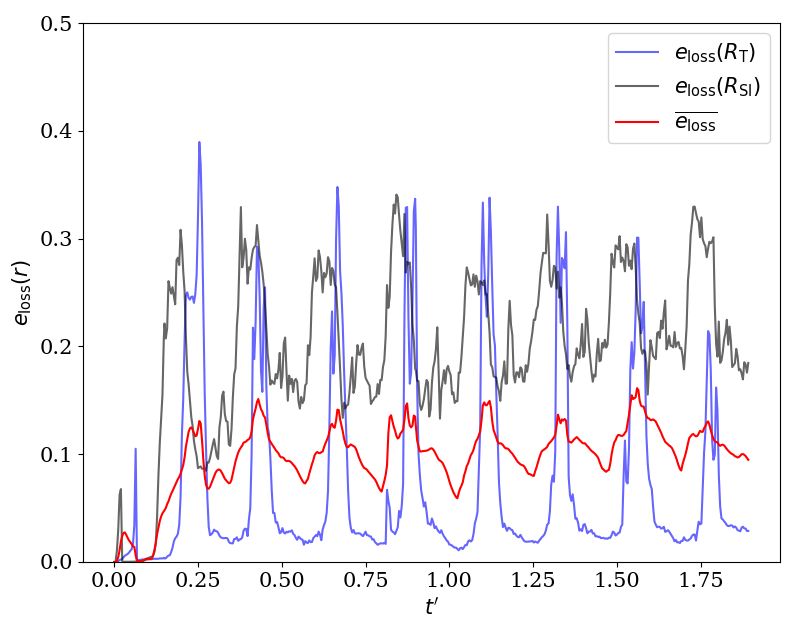}
    \includegraphics[width=\linewidth]{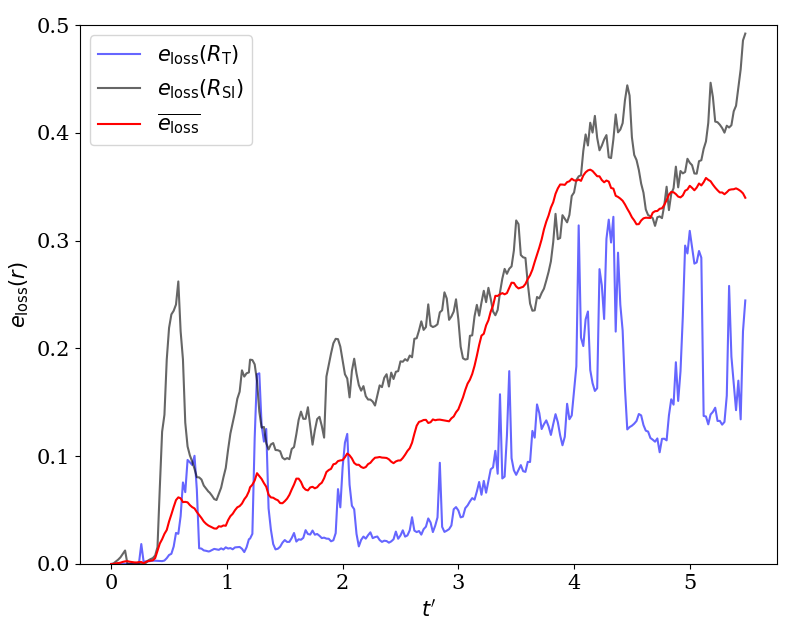}
    \caption{The time evolution of normalized radiative loss $e_{\rm loss}$ (Equation~\ref{eq:eloss}) in RSI20Edd10 (upper panel) and RSI50Edd10 (lower panel). In each panel, the time is in code units. The gray lines are $e_{\rm loss}$ evaluated at the tidal radius $r_{\rm T}=\rp$, blue lines are at the stream-stream collision radius $\rsi$ and the red line is the average value between $\rp$ and $\rsi$.}
    \label{fig:radecirc}
\end{figure}

Figure~\ref{fig:radecirc} compares the radiative loss in RSI20Edd10 and RSI50Edd10. In RSI20Edd10, each stream-stream collision enhances average radiative loss between $\rp=r_{\rm T}$ and $\rsi$. The peaks at $\rsi$ (gray line) are followed by peaks at $\rp$ (blue line), corresponding to the interacting outflow that collides at the orbital plane in between stream-stream collisions. In the absence of stream-stream collision shock and outflow interaction, $e_{\rm loss}$ is low, indicating that shocks are the primary energy dissipation mechanisms in RSI20Edd10. The average $e_{\rm loss}$ between $\rp$ and $\rsi$ (red line) stays around $\sim0.1$ during the repeated collisions. 

In RSI50Edd10, the first three stream-stream collisions lead to the first few peaks of $e_{\rm loss}$ (gray and blue lines). Due to the larger $\rsi$ than RSI20Edd10, the collision strength is reduced, yielding slightly lower radiative loss at both $\rp$ and $\rsi$. After the the outflow subsides and more gas is circularizing, the average $e_{\rm loss}$ between $\rp$ and $\rsi$ (red lines) rises to $\sim0.4$ by the time we stop the simulation. The shocks in the circularizing flow, including the spiral pattern and the ``stream-disk'' shock are the primary sources for radiative loss. 

The $e_{\rm loss}$ evolution in RSI100Edd10 is similar to RSI50Edd10 and also increases during the gas circularization stage. Averaging over time, the stream-stream collision is less efficient in dissipating energy compare to the circularization shocks. When the outflow is too strong, there is little gas accumulation near the black hole to interact with the high eccentricity stream, so the radiative loss is low in between collisions. Once a moderate amount of gas accumulates near the black hole, the circularization shocks can increase $e_{\rm loss}$ rather rapidly. One potentially important caveat is is that our purely hydrodynamics simulations neglect the potential role of magnetic fields in the circularizing flow, which could enhance energy dissipation and angular momentum transport. 

\subsection{Angular Momentum Redistribution}

\begin{figure}
    \centering
    \includegraphics[width=0.95\linewidth]{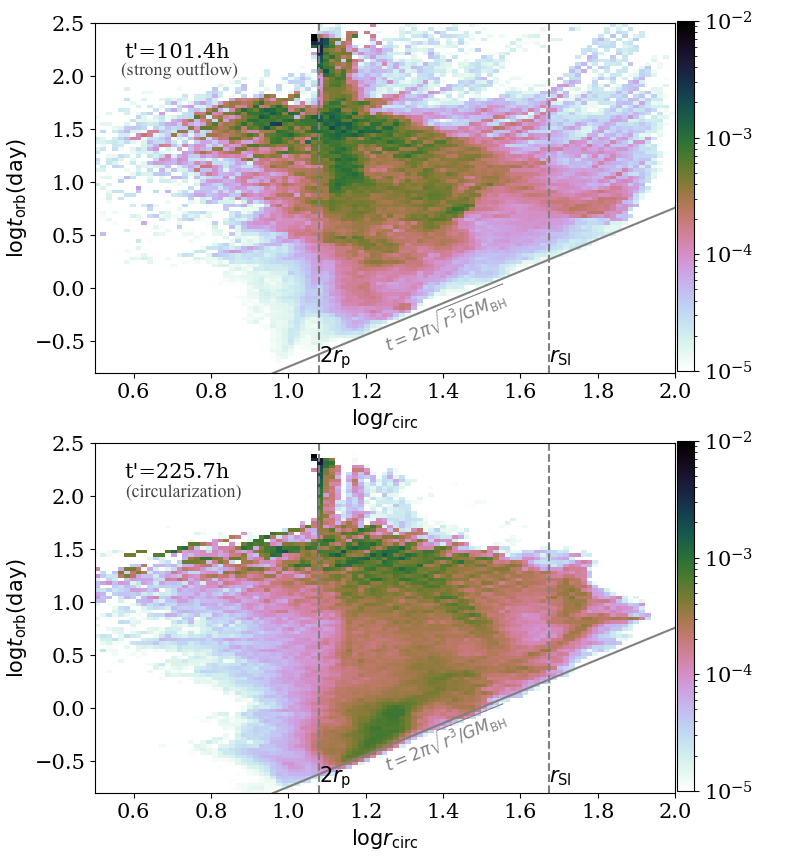}
    \caption{Histograms of Keplerian orbital period $t_{\rm orb}$ and circularization radius $r_{\rm circ}$ for bound gas in RSI50Edd10, the color shows probability for each bin weighted by mass. The upper panel is $t=101.4$h, roughly at the end of strong outflow phase and corresponds to the middle column of Figure~\ref{fig:rsi50edd10_erdens}. The lower panel is $t=225.7$h, when we stopped the simulation and more gas is circularizing. The two gray dashed vertical lines are the circularization orbit of original stellar at $2r_{\rm p}$ and the stream-stream collision radius $\rsi$. The horizontal dashed line is the $t_{\rm orb}$ for a circular orbit at the stellar circularization radius. }
    \label{fig:torb_rcirc}
\end{figure}

For a given time snapshot, we can calculate the gas Keplerian orbital period $t_{\rm orb}$ based on their specific angular momentum $h=r^{3}\psi/(r-1)$ and orbital energy $e=(r^{2}\dot{r}^{2}/(r-1)^{2}+r^{3}\psi^{2}/(r-1))/2.0-GM_{\rm BH}/r$ \citep{tejeda2013accurate}. We approximate gas circularization radius by $r_{\rm circ}=h^{2}/GM_{\rm BH}$. The original stellar orbits has $t_{\rm orb}\approx250$ days and $r_{\rm circ}\approx11\rs$. 

Figure~\ref{fig:torb_rcirc} shows two snapshots of orbital period and circularization radius distribution for all bound gas in strong outflow stage and gas circularization stage. The concentration near $\log t_{\rm orb}\sim2.4$ days, $\log r_{\rm circ}\sim2r_{\rm p}$ corresponds to gas in dense stream with original stellar orbital periods and specific angular momentum. 

The upper panel of Figure~\ref{fig:torb_rcirc} corresponds to the time near the end of strong outflow phase. The stream-stream collision redistributes gas specific angular momentum, so that $r_{\rm circ}$ spans between $\approx 6\rs-\rsi$. A majority of gas has $t_{\rm orb}\approx10-100$ days, which is significantly less than the original stellar $t_{\rm orb}\approx250$ days, but still much longer than the free fall time. The lower panel is when the circularization dominates dissipation. Angular momentum is still redistributed broadly, but as eccentricity drops, more gas acquires orbital properties that are close to circular orbits.
 
In a simplified re-processing picture, the stream-stream collision unbinds part of the marginally-bound stellar debris. The unbound gas propagates outward and assembles the optically-thick reprocessing layer. The more-bound gas will quickly circularize around the black hole near the circularization radius on the free-fall timescale, and eventually be accreted on the viscous timescale. 

In contrast, the simulations indicate that the stream-stream collision disperse gas angular momentum without efficiently damping the orbital eccentricity. So that even if the gas is still bound, they can stay on orbits with relatively high eccentricity and will not return to the black hole after $t_{\rm orb}\sim10^{1-2}$ days.
We conclude that adopting a circularization radius of $2r_{\rm p}$ and free fall timescale may underestimate the initial disk size and the time it takes to circularize bound gas. Furthermore, this bound gas can extend to large radius and will contribute to the reprocessing layer. Hence estimates that relate the mass outflow rate derived from the inferred optical depth of the unbound layer (e.g. $\tau\sim\kappa R \dot{M}_{\rm out}/(4\pi R^{2}v_{\rm out})$) may be skewed by the presence of bound gas on eccentric orbits.

\subsection{Photosphere Properties}\label{subsec:discussion_photosphere}

After the first stream-stream collision, photons are produced via shocks at smaller radius $r\lesssim\rsi$ and diffuse through the optically thick surrounding medium. The relatively constant luminosity (Figure~\ref{fig:mdotlum_mbh}) implies that the diffusion time is longer than the relevant dynamic timescales. The photosphere appearance is determined by both the driving dynamics and the radiation transfer processes.

\begin{figure}
    \centering
    \includegraphics[width=\linewidth]{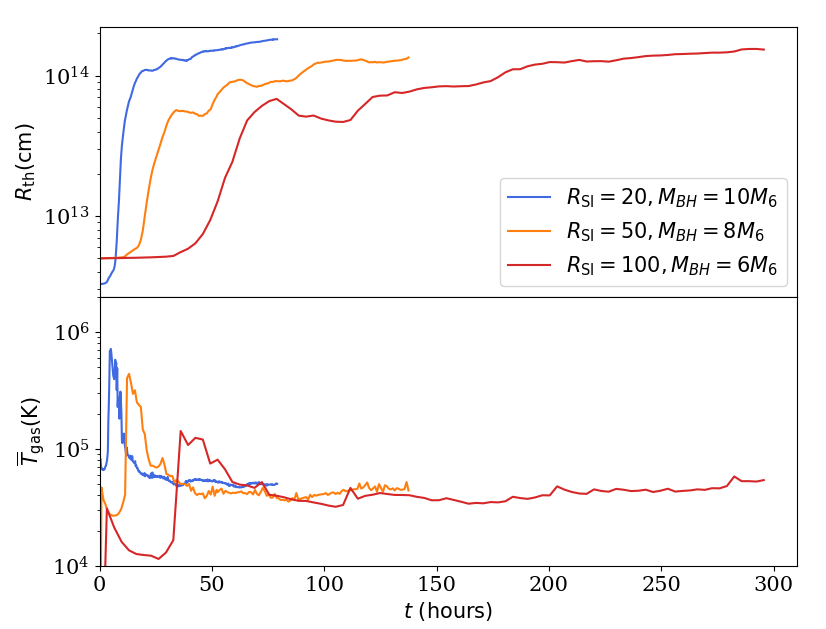}
    \caption{Evolution of thermalization radius $\rth$ (Equation~\ref{eq:rth}, upper panel) and average gas temperature (lower panel) for RSI20Edd10 (blue lines), RSI50Edd10 (orange lines) and RSI100Edd10 (red lines). The prompt rise of $\rth$ and $\overline{T}_{\rm gas}$ is driven by stream-stream collision.  }
    \label{fig:tauavg_mbh}
\end{figure}

We define a thermalization radius $\rth$:
\begin{equation}\label{eq:rth}
    \int_{\rth}^{R_{\rm out}}\sqrt{\kappaa(\kappas+\kappaa)}\rho dr=1
\end{equation}
where $\sqrt{\kappaa(\kappas+\kappaa)}$ is mean opacity from scattering and absorption, $R_{\rm out}$ is the outer boundary radius, where gas density is low and close to floor value. We assume all the lines-of-sight are along radial direction. When $\kappas \ll \kappaa$ this corresponds to the photosphere.
Figure~\ref{fig:tauavg_mbh} shows the evolution of average $\rth$ and the average gas temperature at $\rth$. With this definition of $\rth$, before steam-stream collision, it traces the dense injected stream and roughly follows the average radius of stream orbit. As the initial stream-stream collision launches outflow, $\rth$ promptly increases by about an order of magnitude. Afterward, $\rth$ continues increasing, but with a slower rate and reaches $\sim10^{14}$cm by the time we stop the simulation. In RSI50Edd10 and RSI100Edd10, $\rth$ is comparable to $\rsi$ because gas is primarily circularizing between $r_{\rm ISCO}$ and $\rsi$. In RSI20Edd10, however, $\rth$ primarily traces the outflow moving around the black hole. The thermalization radius expands to $\rth\sim 2\rsi$ with  $\rth \propto t^{0.4}$.

We also compute the time evolution of the average temperature at $\rth$, which peaks $\sim10^{5}-10^6\rm K$ when the gas is heated by the first stream-stream collision and there is very little gas surrounding the stream. Once an outflow develops, the average temperature drops to around a few $10^{4}$K as the average $\rth$ moves outward, forming an optically-thick region extends to $\sim10^{2}\rs$.

In RSI20Edd10, where the optically thick gas flows around the black hole, the distribution of $\rth$ and gas temrpature at $\rth$ are relatively isotropic with view angles. In RSI50Edd10 and RSI100Edd10, as the circularizing flow forms, $\rth$ and $\overline{T}_{\rm gas}$ show greater variation along different lines-of-sight.

\begin{figure}
    \centering
    \includegraphics[width=\linewidth]{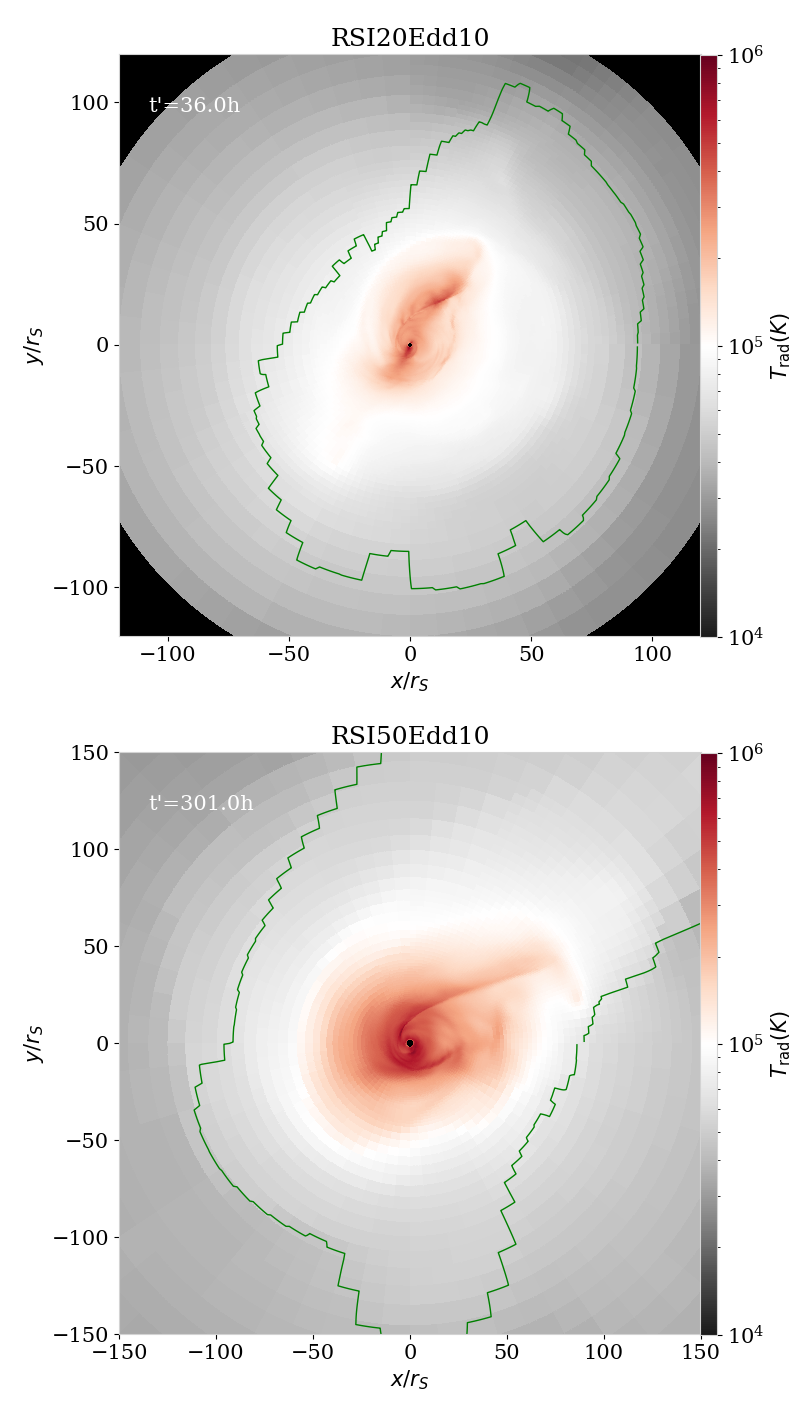}
    \caption{The ``face-on'' radiation temperature snapshot of RSI20Edd10 at $t=36.1$h (the upper panel) and RSI50Edd10 at $t=240$h (the lower panel). The temperature is volume averaged over $\Delta\theta\sim5^{\circ}$ near $\theta=\pi/2$-plane. The green solid line shows $\rth$ along different radial line-of-sight.}
    \label{fig:rsi2050_trad_h}
\end{figure}

\begin{figure}
    \centering
    \includegraphics[width=\linewidth]{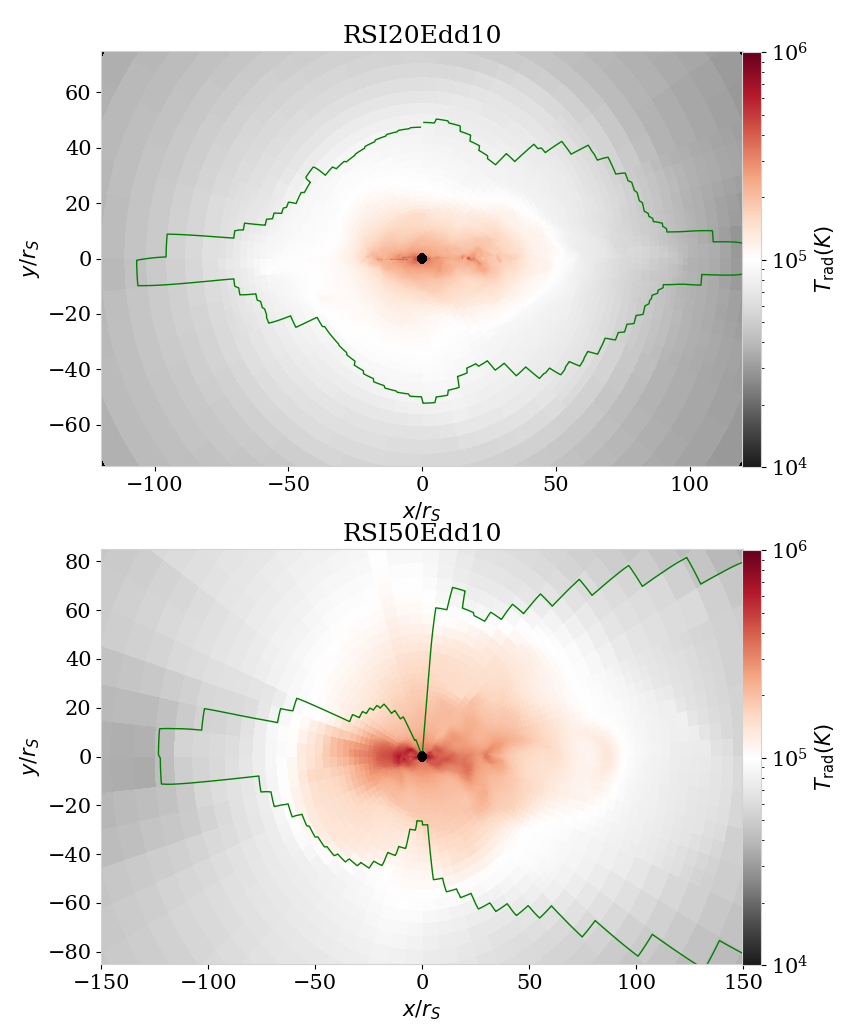}
    \caption{The ``side-view'' radiation temperature snapshot of RSI20Edd10 at $t=36.1$h (the upper panel) and RSI50Edd10 at $t=240$h (the lower panel). The temperature is volume average from $\Delta\phi\sim5^{\circ}$ near $\phi_{\rm inj}$ plane. The green solid line shows $\rth$ along different radial line-of-sight. }
    \label{fig:rsi2050_trad_v}
\end{figure}

In  Figure~\ref{fig:rsi2050_trad_h} and Figure~\ref{fig:rsi2050_trad_v}, we show the radiation temperature $T_{\rm rad}$ and the $\rsi$ distribution in for RSI20Edd10 and RSI50Edd10 together with $\rth$. At this time, RSI20Edd10 corresponds to the middle panel in Figure~\ref{fig:rsi20edd10_repeats}, the stream is reforming and optical-thick gas fills the surrounding region. RSI50Edd10 corresponds to the third column in Figure~\ref{fig:rsi50edd10_erdens}, gas are circularizing near the black hole and flow eccentricity drops. 

The gas circularization in RSI50Edd10 increases the overall density and radiation energy density near the black hole, yielding $T_{\rm rad}\sim10^{6}K$, which is higher than RSI20Edd10 in the central region. Near $\rth$, $T_{\rm rad}$ is roughly $\sim10^{4}$K in both runs in the orbital plane. Perpendicular to the orbital plane, circularizing gas in RSI50Edd10 yields a disk-like geometry and the temperature at $\rth$ varies with viewing angle. In the polar region, gas density is lower and more optically thin, $\rth$ traces gas $\lesssim50\rs$ with temperature of $\sim10^{5}$K. Near the orbital plane, the higher density circularizing gas extends $\rth$ to larger radius, giving $T_{\rm rad}\sim10^{4}$K.

In RSI100Edd10 and RSI50Edd10, we find a similar anisotropy from circularizing gas. The orbital plane is more optically thick than the polar region and $\rth$ is larger. The temperature variation along different radial directions can range one to two orders of magnitude from $\sim10^{4}$K to $\sim10^{6}$K. In comparison, the strong outflow in RSI20Edd10 disperses optically thick gas more uniformly across latitudes and creates a more isotropic photosphere. The anisotropic photosphere may imply potential viewing angle effect on TDE appearance \citep{dai2018unified}. But $\rth$ primarily considers the radially integrated directions, to understand the viewing angle effect on observed spectrum, exploring photosphere properties along more lines-of-sight is of interest for future work.

\subsection{Implication to Pre-peak Multiband Emission}\label{subsec:discussion_multiband}

The estimated average temperature at $\rth$ (Figure~\ref{fig:tauavg_mbh}) will produce optical to UV emission broadly consistent with observations if radiation is nearly thermal. The temperature variation with lattitude, however, suggests there might be different spectral energy distribution (SEDs) when varying lines-of-sight. For example, in Figure~\ref{fig:rsi2050_trad_v}, the residual outflow is optically thick and $\rth\gtrsim100\rs$ on the right hand side of the snapshot. On the left hand side of Figure~\ref{fig:rsi2050_trad_v} (also the last column in Figure~\ref{fig:rsi50edd10_erdens}), gas density is particularly low in the mid-to-high altitude region, and radiation temperature at $\rth$ is $\sim10^{5-6}$K, potentially yielding UV to soft X-ray photons. These regions are likely to be optically thin to scattering, and high energy photons will likely escape more easily. The outflows can also advect out higher energy photons emitted where the flow is hotter. 

The flow geometry is reminiscent of the viewing angle model \citep{dai2018unified,thomsen2022dynamical}, with the important difference that at this time, the accretion flow has not formed  a traditional ``hot coronae'' or a super-Eddington disk. Instead, the high energy photons are primarily from shocks in the inner circularization region. In such a flow geometry and velocity field, X-ray emission may emerge in TDE systems before a classical geometrically and optically thick super-Eddington disk forms.

Figure~\ref{fig:unbound_mbh} suggests that $\sim 10-20\%$ of outflow launched during the first few collisions is unbound. If we integrate unbound $L_{\rm KE}$ in Figure~\ref{fig:unbound_mbh} for times when $\dot{M}_{\rm out, ub}>0.1\Medd$ and $L_{\rm KE}>10^{42}\rm erg~s^{-1}$, the total unbound kinetic energy through the spherical shell at $150\rs$ is $E_{\rm KE}\approx1.2\times10^{49}\rm erg$ in RSI50Edd10, and the total unbound mass is $M_{\rm ej,ub}\approx1.1\times10^{-3}M_{\odot}$. With similar estimation, we have total unbound kinetic energy $E_{\rm KE}\approx6.8\times10^{48}\rm erg,~1.9\times10^{49}\rm erg$, total unbound mass $M_{\rm ej,ub}\approx7\times10^{-4}M_{\odot},~1.7\times10^{-3}M_{\odot}$ in RSI100Edd10 and RSI20Edd10. If the unbound outflow can escape without significant energy dissipation, their collision with circum-nuclear medium may be relevant to producing the synchrotron emission at larger radius.

\citet{goodwin2023radio} found that the early radio emission in AT2020vwl is likely related to sub-relativistic outflow launched before the optical peak. For AT2020vwl, the estimated black hole mass is smaller than the adopted $M_{\rm BH}$ in this work. Considering for the emission radius $\sim10^{16}\rm cm$ (unbound outflow travel distance), reducing $M_{\rm BH}$ but keepin a similar $\beta=1$ will lead to a larger collision radius $\rsi\sim10^{3}\rs$ and weaker collision. Higher $\beta\approx4$ can give $\rsi\sim50\rs$, which is similar to RSI50Edd10. The different black hole mass and unknown $\beta$ introduce uncertainties whhen extrapolating our simulation results. Nevertheless, estimating the unbound gas properties from simulation near $\sim10^{16}\rm cm$ provides an interesting comparison.

\begin{figure}
    \centering
    \includegraphics[width=\linewidth]{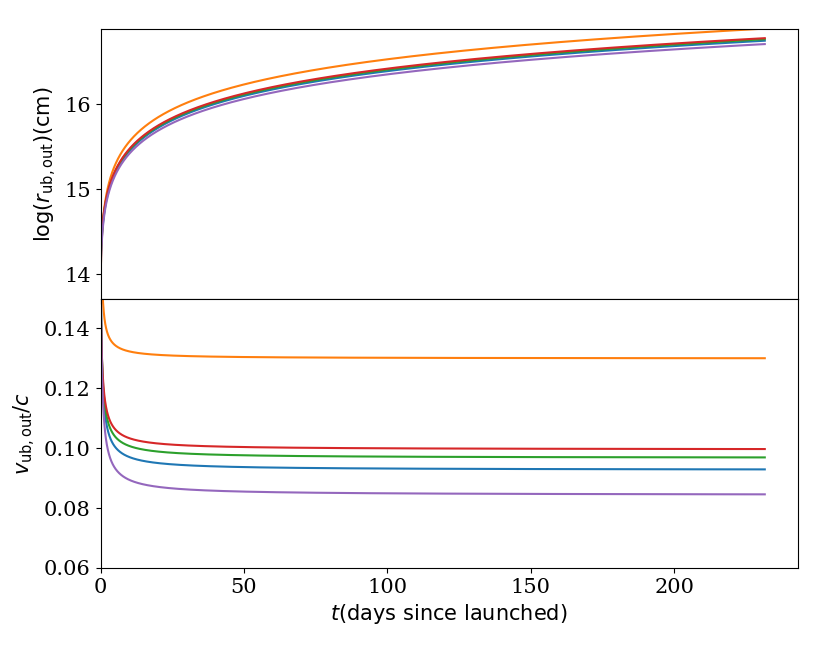}
    \caption{Inferred unbound outflow travel distance from the black hole (the upper row) and total velocity magnitude (the lower panel) in RSI50Edd10. We randomly sample velocity and position from five points in Figure~\ref{fig:rsi50edd10_unbound} to represent gas parcels in the unbound outflow when they are just launched. Then we integrate the ballistic trajectories to extrapolate their travel distances and velocities, and infer the sub-relativistic outflow associated with potential radio emission. The blue, orange, green, red and purple lines corresponds to sample points with initial $(r,~\theta)=(35.08,0.78),~(17.91,1.05),~(40.05,1.48),~(49.80,1.48),~(24.95,2.26)$ on the $\phi=1.66\pi$ plane in Figure~\ref{fig:rsi50edd10_unbound}.}
    \label{fig:unbound_radio}
\end{figure}

Here we use RSI50Edd10 as an example to extrapolate unbound outflow properties at $\sim10^{16}\rm cm$, where the radio emission is estimated to be produced. The total unbound kinetic energy is $\approx10^{49}\rm erg$ at $150\rs$. To reach the emission radius $\sim10^{16}\rm cm$, the outflow need to travel to $\approx4000\rs$. The kinetic energy change between $150\rs$ and $\sim4000\rs$ is roughly within order of magnitude, giving approximately consistent total kinetic energy. 

To estimate the outflow velocity and travel distance beyond the simulation domain, we randomly sample five points from unbound outflow in Figure~\ref{fig:rsi50edd10_unbound}, about $12$ hours after they are launched by the second stream-stream collision in RSI50Edd10. By integrating the ballistic trajectories using the sampled points as initial condition, we show their distance to the black hole and velocities in Figure~\ref{fig:unbound_radio}. The sampled unbound outflow travels to $\sim10^{16-17}\rm cm$ after 100-200 days since launched, with velocity $\sim0.1c$. Comparing with \citet{goodwin2022at2019azh}, the selected unbound outflow in RSI50Edd10 seem to be best-fitted by the assumption of electron energy fraction $\epsilon_{e}=0.1$, with slight preference of conical geometry. The exact comparison requires a lower black hole mass and time-dependent mass fallback rate. Unbound outflow along different lines-of-sight might show distinct properties too. These factors will be considered in the future work.

We can also apply a similar extrapolation based on ballistic trajectories to bound outflow. The majority of outflow in RSI50Edd10 is bound and may return to the black hole on the timescale of their orbital period. We also randomly sample five points from bound gas on the same plane as the above unbound gas snapshot, which is not shown in Figure~\ref{fig:rsi50edd10_unbound}. We found they will return to the black hole on timescales of $10-100$ days, and can travel as far as $\sim10^{15}\rm cm$. The timescale and distance is generally order-of-magnitude consistent with Figure~\ref{fig:torb_rcirc}. If some of the earliest-launched outflow gas survives traveling through the surrounding medium, it will return to the black hole with velocity $\sim0.1c-0.2c$ after $10-100$ day, and a super-Eddington disk may already form from the circularizing gas. The interaction between the returning outflow and the accretion disk or disk wind may also produce multiband emission.

\subsection{Comparison with Previous Works}\label{subsec:discussion_previouswork}

The repeating interactions in RSI20Edd10 provide a limit of strong stream-stream collision when $\rsi$ is small and $\mdot$ is high. Dropping $M_{\rm BH}$ in RSI50Edd10 and RSI100Edd10, $\rsi$ moves away from the black hole. The debris stream in both of simulations show multiple collisions before settling to circularization. However, only the first few collisions drive strong outflow before the returning stream disperses near pericenter. In the limit of low black hole mass, $\rsi$ is likely to be near the apocenter instead of near the pericenter. In this scenario, the returning stream might be more expanded compare to small $\rsi$, and the apocenter shock between the streams might be the primary shock dissipation mechanism \citep{shiokawa2015general, piran2015disk, krolik2016asassn}. With non-zero spin and large penetration factor $\beta$, \citet{andalman2022tidal} also finds multiple stream-stream collisions before accretion disk formation. Understanding how different $\rsi$ and debris stream orbits lead to different stream-stream collision and their effects on outflow rate and emission source will be interesting future works.

In the simulations, we found that various shocks are the leading pre-peak emission mechanism instead of accretion, consistent with recent studies \citep{lu2020self, steinberg2022origins, ryu2023shocks}. The simulations presented in this paper does not include magnetic fields, which are essential to excite magnetic rotational instability (MRI) \citep{balbus1991powerful}. MRI may dominate angular momentum transport and facilitate more efficient accretion, and possibly lead to jet formation \citep{shiokawa2015general, andalman2022tidal, curd2023strongly}. Our results are also not likely to be a good approximation if the black hole spin is large. The Lense-Thirring precession induced by spinning black hole can introduce vertical offsets between the fallback and returning stream \citep{hayasaki2016circularization,jiang2016prompt,batra2023general,jankovic2023spininduced} and delay stream self interaction \citep{guillochon2015dark, hayasaki2016circularization}. General relativistic effects may also affect the post-collision circularization process and accretion disk formation \citep[see e.g.]{sadowski2016magnetohydrodynamical, liptai2019disc, curd2021global}. 

The circularizing flow formed in RSI50Edd10 and RSI100Edd10 is moderately eccentric. Lacking magnetic field, the truncated spiral shock near the black hole ISCO is likely to be the primary angular momentum transport mechanism. Similar spiral patterns are identified in previous work \citep[see e.g.]{bonnerot2021formation,curd2021global,ryu2023shocks}. The spiral shock is truncated by the dense fallback stream in our simulations. If the circularizing flow eventually reaches an average density that is comparable to the fallback stream, the spiral shock may be able to extend to larger radius and further facilitate accretion. \citet{curd2023strongly} studies the scenario where a debris stream running into a magnetically dominated accretion disk. They find that for a range of density contrast between the stream and disk, hydrodynamical shocks is more important to the angular momentum transport than magnetic stress. However, it seems no significant spiral structure forms globally over the disk. Understanding the relative importance between the hydrodynamical shock, magnetic stress and radiation viscosity will help understand the angular momentum transport mechanism in TDEs accretion.

The circularizing flow shows noticeable eccentricity in the vertical direction. Due to the interaction between thickened returning stream and fallback stream, the side that is closer to stream-stream collision has larger scale height ($H/R\sim\rm tan(\pi/4)$) than the side closer to pericenter ($H/R\sim\rm tan(\pi/8)$). Eccentric TDE disks are discussed by previous work assuming inefficient circularization. The models qualitatively agrees on the eccentric geometry of small (large) scale height at pericenter (apocenter), with different primary dissipation mechanisms \citep{piran2015disk, liu2017disc, zanazzi2020eccentric}. In our simulations, the different scale height near the pericenter side and stream-stream collision side is more related to the outflow instead of compression by nozzle shock. Understanding the thermal emission from eccentric disk and their difference to near-Keplerian disk can bring new insights to TDE emission. When including magnetic field, understanding the competing effect of angular momentum transport and orbital energy dissipation by MRI and stress tensors in eccentric TDE disk may also be relevant to shape the long-term accretion physics \citep{chan2018magnetorotational,chan2022nonlinear} and jet launching processes.

\section{Conclusions}\label{sec:conclusion}

We find that stream-stream collision can delay gas circularization by launching strong outflows and depleting gas near the black hole. For fallback rates of $10\Medd$, the instantaneous outflow rate can reach $\mdotout\approx5\Medd-10\Medd$ depending on the black hole mass. The accretion rate onto the black hole rarely exceeds Eddington value. The stream-stream collision is likely to happen multiple times until the returning stream expands significantly after the pericenter passage. The outflow subsides when the returning stream disperses near the pericenter, and gas accumulation and circularization only starts after the outflow rate drops. 

For black hole masses $6-10M_{6}$ and mass fallback rate $10\Medd$, the pre-peak emission in TDEs is primarily powered by various shocks instead of direct accretion. The initial stream-stream collision can produce prompt emission $\sim10^{44}\rm erg~s^{-1}$ (Figure~\ref{fig:mdotlum_mbh}).
When the system in in the strong outflow stage, the collision shock and shocks produced by outflow interaction are the major emission sources. After the outflow rate decreases and gas start to circularize, the accretion flow develops multiple circularization shocks to power the emission, including the ``stream-disk'' shock between the dense stream and circularizing gas (Section~\ref{subsec:discussion_emissionsource}) and truncated spiral shocks near the ISCO.

The radiation efficiency $L/\dot{M}c^{2}$ is time-dependent, with typical values $>10\%$ during the strong outflow stage, and drops to $\sim5\%$ in gas circularization stage (Figure~\ref{fig:lumeff_mbh}). The luminosity does not tightly follow the accretion rate, especially when the outflow rate is high because the shocks, not accretion, dominate emission. Despite the super-Eddington fallback rate, the bolometric luminosity is sub-Eddington for the black hole mass.

The outflows produced by the stream-stream collisions are optically thick. As material propagates away from the black hole, the average $\rth$ increases to the order of $\sim100\rs$, and we estimate the photosphere expands to the order of $\sim10^{14}\rm cm$. The hot post-shock gas at $<\rsi$ has typical temperature $\sim10^{5}-10^{6}$K, but the average gas temperature near $\rth$ drops to $\sim10^{4}$K (Figure~\ref{fig:tauavg_mbh}). The photosphere can be highly aspherical, especially when circularization concentrates more gas near the orbital plane. The photosphere size and temperature may show orders of magnitude variation (Figure~\ref{fig:rsi2050_trad_h}, Figure~\ref{fig:rsi2050_trad_v}), potentially leading to a strong viewing angle dependence for the  SED even before a super-Eddington disk forms.

When the fallback and returning stream are both thin and maintain sufficient density contrast with surrounding gas, $\sim 1-2\Medd$ of outflow produced in stream-stream collision is unbound. The average radial velocity is at the order of $0.1c$, slightly exceeds the local escape velocity (Figure~\ref{fig:unbound_mbh}). The integrated kinetic energy through spherical shells $\gtrsim 150\rs$ carried by the unbound gas reaches $\sim10^{49}\rm erg$. For smaller collision radius (larger black hole mass for the same $\beta$, or larger $\beta$ for the same black hole mass) and higher fallback rate, the collisions tend to drive higher unbound mass flux and kinetic luminosity. After the strong outflow stage, the circularization stage rarely produces large amounts of unbound gas in these simulation, but bound gas may still reach large radius before falling back to the inner flow region.

When studying TDE systems, including bound-bound and bound-free opacity is essential to capture the optical band emission. The classical Kramer's law type of free-free opacity can be a few orders of magnitude smaller than the opacity functions including the atomic process (for example, OPAL and TOPs opacity). During circularization stage, some gas in the post-shock flow can be heated to $\sim10^{6}$K close to the black hole. There might be directions where gas density is low and therefore optically thin to these high energy photons, for example, near the polar region and away from the stream-stream collision site. These results suggest it is possible to produce soft X-ray band emission prior to the optical peak.

\begin{acknowledgments}
We would like to thank Phil Chang and Maria Renee Meza for supportive collaboration. We are grateful to stimulating discussions and helpful insights from Kate Alexander, Ryan Chornock, Brenna Mockler, Ann Zalbludoff, Daichi Tsuna, Wenbin Lu, Enrico Ramirez-Ruiz, Jane Lixin Dai, Cl\'{e}mont Bonnerot, Phil Hopkins. XH is supported by the Sherman Fairchild Postdoctoral Fellowship at the California Institute of Technology. This work used Stampede 2 at Texas Advanced Computing Center through allocation AST150042 from the Advanced Cyberinfrastructure Coordination Ecosystem: Services $\&$ Support (ACCESS) program, which is supported by National Science Foundation grants 2138259, 2138286, 2138307, 2137603, and 2138296. Resources supporting this work were also provided by the NASA High-End Computing (HEC) Program through the NASA Advanced Supercomputing (NAS) Division at Ames Research Center. Support for this work was provided by the National Aeronautics and Space Administration under TCAN grant 80NSSC21K0496 and the National Science Foundation under grant 2307886. The Center for Computational Astrophysics at the Flatiron Institute is supported by the Simons Foundation. 
\end{acknowledgments}

\bibliography{ref}

\appendix
\section{Stream Vertical Structure Near Pericenter}\label{appendix:resolution}
\begin{figure}
    \centering
    \includegraphics[width=0.6\linewidth]{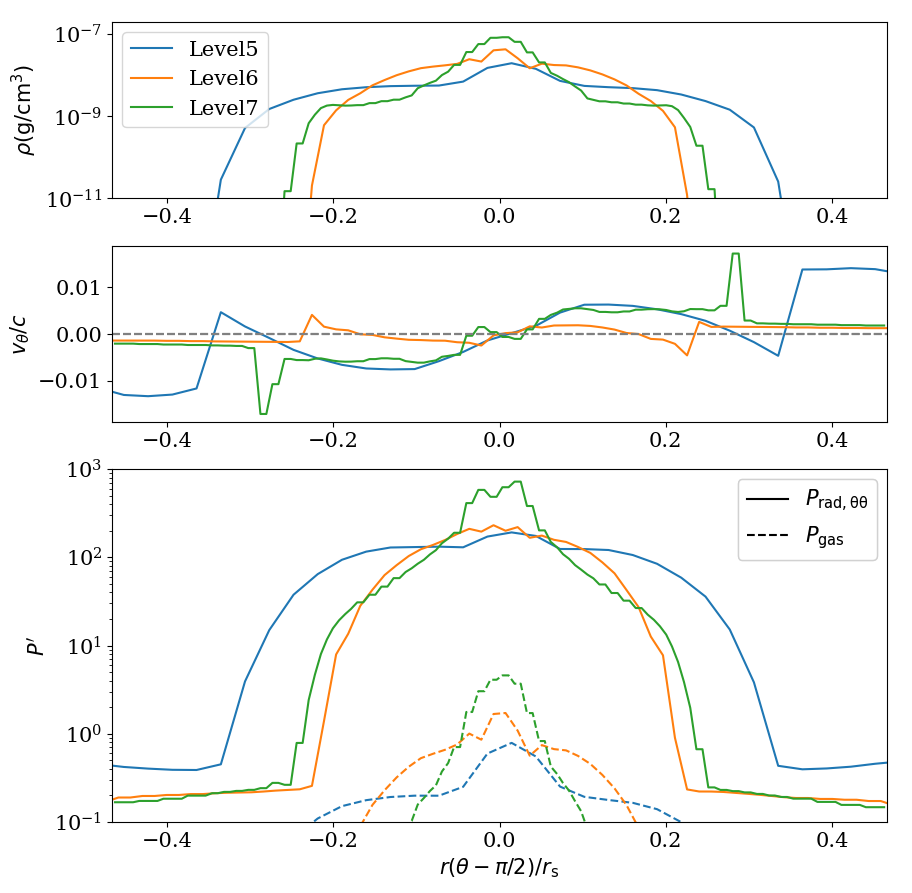}
    \caption{The vertical stream density (the first row), $\theta$ direction velocity (the second row) and pressure (the third row) at $r=4.75\rs$ (the gray circle in Figure~\ref{fig:resedd10_erdens}), $t=26.2$h. The blue, orange and green lines corresponds to level five, six and seven. In the third row, the solids lines are the $\theta\theta$ component of radiation pressure tensor, the dashed lines are gas pressure. }
    \label{fig:resedd10_circlecut}
\end{figure}

Figure~\ref{fig:resedd10_circlecut} shows the vertical density profile of the streams for different resolutions discussed in Section~\ref{subsec:result_resolution}. Increasing resolution leads to more compression and higher central density. At this time, the stream is expanding vertically, with typical velocity $v_{\theta}\lesssim 0.01 c$. We compare the radiation pressure $P_{\rm rad,\theta\theta}$ and gas pressure $P_{\rm gas}$ and find the stream is radiation pressure dominated. The expansion velocity from level five to seven does not follow a linear trend, suggesting there might be unresolved physics. It is also likely that the stream evolves to slightly different dynamical stages when changing resolution, resulting different velocity profiles.

Compared to recent studies focusing on the pericenter compression \citep{bonnerot2022nozzle, bonnerot2022pericenter, coughlin2023dynamics}, the central density in Figure~\ref{fig:resedd10_circlecut} is few orders of magnitude lower. \citet{bonnerot2022nozzle} proposes that the compression is near adiabatic until a shock forms near the orbital plane. As the shock propagates outward, it quickly re-expands the stream to roughly the width before pericenter passage. The stream optical depth is sufficently large so that losses due to radiative diffusion might be negligible. 

\begin{figure}
    \centering
    \includegraphics[width=0.6\linewidth]{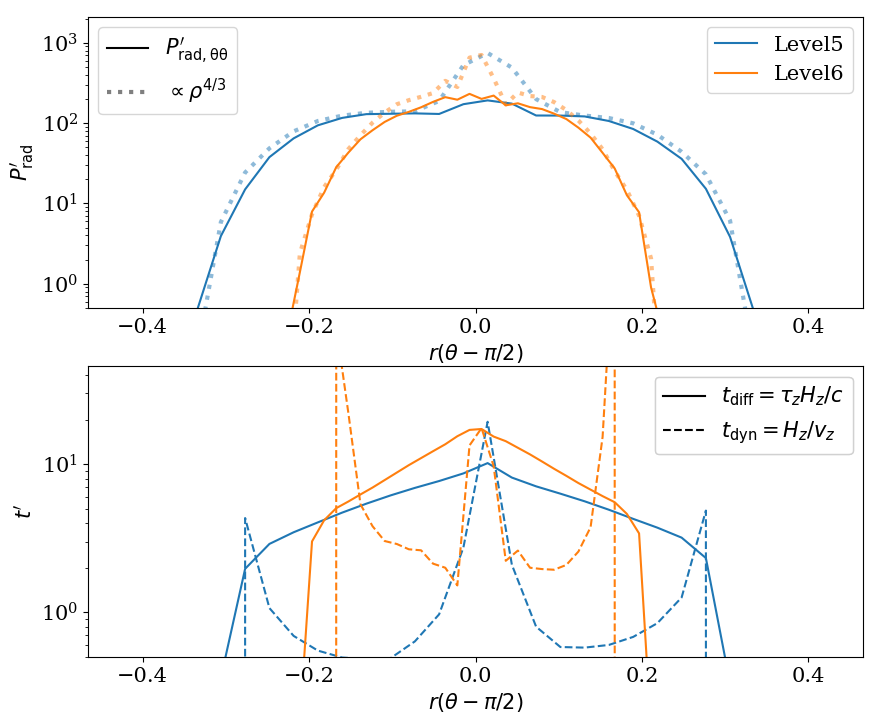}
    \caption{The vertical stream pressure (the upper panel) and estimated timescales (the lower panel) at $r=4.75\rs$ (the gray circle in Figure~\ref{fig:resedd10_erdens}), $t=13.1$h for level five (blue lines) and level six (orange lines). In the upper panel, the solid line are the $\theta\theta$ component of radiation pressure tensor. The dotted lines are fitted curve that scales as $\propto\rho^{4/3}$. In the lower panel, the solid lines are estimated vertical diffusion timescale, and the dashed lines are estimated vertical dynamical timescale.}
    \label{fig:resedd10_tdiff}
\end{figure}

Despite the different approaches, we find roughly consistent large optical depth across the stream as in \citet{bonnerot2022nozzle}. We estimate the vertical dynamical time $t_{\rm diff}=H_{\rm z}/v_{\rm z}\approx r\Delta\theta/v_{\theta}$, where $r\Delta\theta$ and $v_{\theta}$ are distance to the orbital plane and expansion velocity in the $\theta$ direction (positive corresponds to expansion). The vertical diffusion timescale can be estimated as $t_{\rm diff}=\tau_{\rm z}H_{\rm z}/c\approx\kappa\rho (r\Delta\theta)^{2}/c$, where $\kappa=(\kappas+\kappa_{\rm a})$ is the total opacity from scattering and absorption. Comparing these two timescales, the radiation diffusion is relatively slow in the most of region within the stream, except for central region with highest density. The radiation pressure roughly follows the adiabatic prediction $\propto\rho^{4/3}$ in the regions where the diffusion timescale is longer than the dynamical timescale, except for the central region. The radiation pressure is also enhanced compared to adiabatic assumption in the central region, where the most of dissipation happens. 

One of the differences with \citet{bonnerot2022nozzle} is that we do not capture the shock launched near the orbital plane that rebounds the stream vertically. Hence, the velocity near the mid-plane is potentially lower than the velocity that would be driven by the rebounding shock, resulting in a slower dynamical time. The difference could related to the lower resolution in our global simulations. To understand the details of the role of advective and diffusive radiation flux, a localized high resolution study is required in the future work. 

\citet{steinberg2022origins} present a series localized pericenter studies in addition to their global simulation, finding that the typical global resolution cannot resolve the pericenter shock, yielding an artificially large stream scale height, but having limited effect on longer term global evolution. However, including the heating injected by recombination can significantly inflate the stream after pericenter passage, leading to a more expanded stream. \citet{coughlin2023dynamics} also propose that the recombination may increase stream entropy near the pericenter passage. Both works suggest that the dissipation associated with recombination is unlikely to drive a significant transient in early time. An earlier study \citet{kasen2010optical} suggested that it may play an important role to generate extra radiation during the nozzle shock. Understanding this effect and its interplay with radiation is an interesting topic for future work, but would require more localized simulations focused in the pericenter dynamics.

In this work, we fix the $\mdot=10\Medd$ when studying the resolution effect. For future work with a time-dependent $\mdot$, resolving the pericenter optical depth might be more challenging. Since the optical depth $\tau\propto\rho\propto\mdot$, more resolution is likely required to avoid over-cooling  when the $\mdot$ is sub Eddington at the earliest time when fallback commences. More aggressive use of AMR may be needed to resolve this in the future.

\end{CJK*}
\end{document}